\documentclass[%
reprint,
 amsmath,amssymb,
 aps,
pra,superscriptaddress
]{revtex4-1}

\usepackage{graphicx}% Include figure files
\usepackage[caption=false]{subfig}
\usepackage{pgfplots}
\usepackage{soul}
\usepackage{physics}
\usepackage{stackengine}
\usepackage{svg}
\DeclareFontFamily{OT1}{pzc}{}
\DeclareFontShape{OT1}{pzc}{m}{it}{<-> s * [1.10] pzcmi7t}{}
\DeclareMathAlphabet{\mathpzc}{OT1}{pzc}{m}{it}
\usepackage{dcolumn}% Align table columns on decimal point
\usepackage{bm}% bold math
\usepackage{hyperref}% add hypertext capabilities
\usepackage{xcolor}
\usepackage{mathtools}
\usepackage{tikz}
\usetikzlibrary{patterns}
\usepackage{soul}
\newcommand\underrel[3][]{\mathrel{\mathop{#3}\limits_{%
      \ifx c#1\relax\mathclap{#2}\else#2\fi}}}
\usepackage{natbib}
\usepackage{braket}

\newcommand{\qg}{\mathbf{q}}

\newcommand{\kg}{\mathbf{k}}

\begin{document}

%\preprint{APS/123-QED}

\title{Evolution of the unitary Bose gas for broad to narrow Feshbach resonances}% Force line breaks with \\
%\thanks{A footnote to the article title}%

\author{D.J.M. Ahmed-Braun}
 \email [Corresponding author:  ]{d.j.m.braun@tue.nl}
\affiliation{Eindhoven University of Technology, P.O. Box 513, 5600 MB Eindhoven, The Netherlands}
\author{S. Musolino}
\affiliation{Eindhoven University of Technology, P.O. Box 513, 5600 MB Eindhoven, The Netherlands}
\author{V. E. Colussi}
\affiliation{INO-CNR BEC Center and Dipartimento di Fisica, Universit\`a di Trento, 38123 Povo, Italy}
\author{S. J. J. M. F. Kokkelmans}
\affiliation{Eindhoven University of Technology, P.O. Box 513, 5600 MB Eindhoven, The Netherlands}

\date{\today}% It is always \today, today,
             %  but any date may be explicitly specified

\begin{abstract}
We study the post-quench dynamics of unitary Bose gases using a two-channel model, focusing on the effect of variations in the width of the Feshbach resonance due to density changes. We generally find that increasing the density leads to a corresponding increase in the production of closed channel molecules, a decrease in the build up of quantum depletion and a transition from linear to quadratic early-time growth of the two-body contact as well as the condensed pair fraction. Motivated by the presence of closed-channel molecules in the unitary regime, we study the embedded two-body problem finding a transition from open to closed-channel dominated dimers due to many-body effects.
\end{abstract}

\pacs{Valid PACS appear here}% PACS, the Physics and Astronomy
                             % Classification Scheme.
%\keywords{Suggested keywords}%Use showkeys class option if keyword
                              %display desired
\maketitle

%\tableofcontents
%\linenumbers
\section{\label{sec:level1}Introduction}
The magnetic tunability of Feshbach resonances \cite{FESHBACH1958357} in ultracold gases makes it possible to experimentally control the two-particle interaction strength as characterized by the s-wave scattering length $a(B)$
\begin{equation}
\label{eq:ScatteringFeshbach}
a(B) = a_{\mathrm{bg}}\left(1-\frac{\Delta B}{B-B_0}
\right),
\end{equation} 
with background scattering length $a_{\mathrm{bg}}$, magnetic resonance width $\Delta B$ and resonance position $B_0$ \cite{art:chin}. 
By diabatically quenching the scattering length to unitarity ($\abs{a(B)}\rightarrow \infty $) \cite{art:makotyn,art:klauss,art:eigen,art:eigenpretherm} , it is possible to beat disastrous per-particle losses in bosonic gases that scale as $n^2a^4$, with atomic density $n$.  At unitarity the density scales remain finite and according to the ``Universality Hypothesis'' all properties of unitary quantum gases should scale continuously with the Fermi scales $k_{\mathrm{n}} = (6\pi^2 n)^{1/3}$,  $E_{\mathrm{n}} = \hbar^2 k_{\mathrm{n}}^2/(2m)$ and $t_{\mathrm{n}} = \hbar/E_{\mathrm{n}}$, where $m$ is the atomic mass \cite{art:ho}.
\typeout{the scattering length effectively drops out of the set of length scales, such that only the density length scale $n^{-1/3}$ remains. 
This is the foundation of the ``Universality Hypothesis'', which considers the full characterization of system properties in terms of density scales, or Fermi scales, including the Fermi wave number $k_{\mathrm{n}} = (6\pi^2 n)^{1/3}$, energy $E_{\mathrm{n}} = \hbar^2 k_{\mathrm{n}}^2/(2m)$ and time $t_{\mathrm{n}} = \hbar/E_{\mathrm{n}}$, where $m$ is the atomic mass \cite{art:ho}.} This universality and the associated scale invariance relates the properties of ultracold gases at unitary to other seemingly unrelated strongly correlated systems, such as the quark-gluon plasma and the inner crust of neutron stars \cite{0034-4885-72-12-126001,PhysRevLett.91.102002,book:Castin2012}. 
However, the non-universal scales associated with the Feshbach resonance in addition to the finite size of Efimov states in a unitary Bose gases \cite{Efimov1,Efimov2,art:klauss,art:efimovquench,art:colussi_three} can alter the universal scaling and time-dependence of system properties. \par 
To describe the interplay between resonance and density scales, the vacuum classification of a Feshbach resonance must be revisited in the many-body context \cite{PhysRevLett.108.250401}.
The magnetic resonance width is momentum dependent and equal to
\begin{align}
\label{eq:DeltaB}
\Delta B  =  \frac{\hbar^2 k}{\delta \mu  m R^*},
\end{align}
where $\delta \mu$ is the difference in the magnetic moment between two free atoms and the Feshbach molecule and $R^*$ is the resonance width parameter.
In the vacuum classification scheme, the momentum scale $k \sim 1/a_{\mathrm{bg}}$ in estimating $\Delta B$, such that $\abs{R^*/a_{\mathrm{bg}}} \ll 1$ corresponds to a broad resonance. In the many-body classification scheme on the other hand, the momentum $k$ is set by $k_{\mathrm{n}}$ \cite{PhysRevLett.108.250401}, such that $k_n R^* \ll 1 $ corresponds to a broad resonance. Furthermore, whereas the vacuum classification is fixed by the specifics of the resonance and the atomic species, the many-body classification varies with the density of the gas. 
\par 
In this work, using the many-body classification of the resonance width, we go beyond the single-channel models of Refs. \cite{art:MusolinoPairFormation,art:colussimk,art:sykes,art:colussi_three,art:efimovquench,PhysRevA.90.063626,PhysRevA.91.013616,art:Ancilotto2015,PhysRevA.93.033653,heras2018early,PhysRevLett.124.040403,PhysRevA.102.063314,musolino2021boseeinstein} and study the post-quench evolution of the unitary Bose gas using a two-channel model that explicitly contains the multichannel nature of the Feshbach resonance. The one and two-body correlation dynamics in the system are modelled using the cumulant method \cite{art:MusolinoPairFormation,art:colussimk}.
The correlation dynamics are studied over a range of $k_{\mathrm{n}} R^*$, where we characterize the onset of non-universal effects in the dynamics of the two-body contact, the atomic, pair and molecular condensate fractions and the quantum depletion. To describe the pair formation, the single-channel formalism of Ref. \cite{musolino2021boseeinstein} is generalized to the two-channel model. Here, the presence of the molecular fraction in the two-channel system provides us with an additional probe of the many-body state at unitarity by using molecular spectroscopy \cite{PhysRevLett.95.020404}. 
To explain the presence of molecules in the unitary regime, we study the impact of many-body effects on the embedded few-body problem. This problem is a generalization of the single-channel formalism of Ref. \cite{art:colussimk}, which considers both the dressing due to the channel couplings and the impact of the quantum statistics of the medium in the spirit of the Cooper pair problem \cite{book:Zwerger,art:nozieres,art:holland}.
We observe that, for all considered values of $k_{\mathrm{n}} R^*$, the embedded dimers become the dominant contribution to the quantum depletion as the system evolves in the unitary regime.
\par
The paper is outlined as follows. In Sec.\;\ref{subsec:CoupledChannelsTmatrix} we review the two-channel model of a Feshbach resonance, which provides a foundation for the many-body model that follows in Sec.\;\ref{subsec:manybodyeq}. We then apply the cumulant expansion to include up to two-body correlations in the dynamics. Within this formalism we derive the expressions for the pair condensate fraction and the dynamical two-body contact. Next, we proceed with the formalism of the embedded two body problem. These concepts are subsequently used to understand the results presented in Sec.\;\ref{sec:comparison} and to form the conclusions in Sec.\;\ref{sec:conclusion}.
\section{\label{sec:coupledchannelsmodel} Two-channel model}
In this paper we aim to study the onset of non-universal effects over a range of resonance widths $k_{\mathrm{n}}R^*$. To this extend we develop a two-channel model that captures the multichannel nature of Feshbach resonances. We start by reviewing the two-channel model in vacuum in Sec.\;\ref{subsec:CoupledChannelsTmatrix} and then solve the many-body version in Sec.\;\ref{subsec:manybodyeq} using the method of cumulants. Here we derive the two-channel generalizations of the pair condensate fraction and the two-body contact. In Sec.\;\ref{subsec:Embeddedtwobodyinteractions} we include many-body effects in the vacuum two-channel model in an effort to investigate how the few-body physics is altered by the medium.

\subsection{\label{subsec:CoupledChannelsTmatrix} The two-body problem}
We employ a two-channel model in order to describe the effect of the resonance width, classified in terms of $k_{\mathrm{n}}R^*$, on the dynamics of the quenched unitary Bose gas. To that end we begin with a brief review of Feshbach resonance theory \cite{book:Feshbach} which serves as a basis for the generalizations to the many-body context in what follows. \par
\begin{figure}[t!]
\centering
\includegraphics[width=8.6cm,height = 5.5cm]{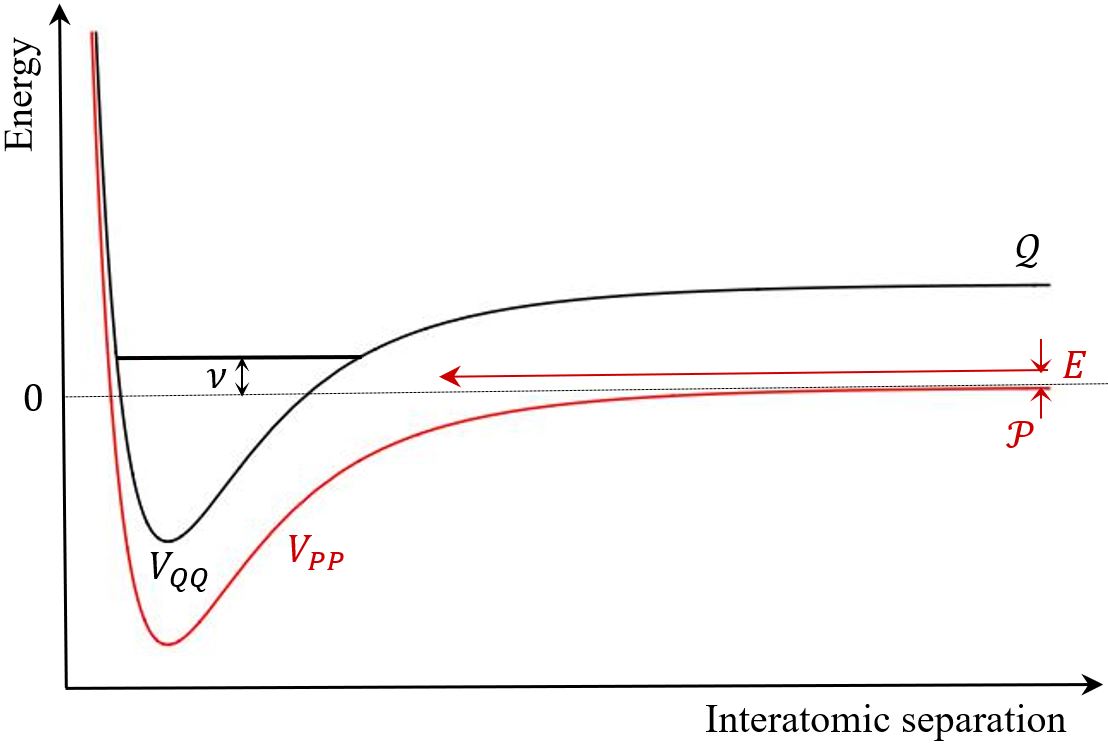}
\caption{Schematic of the basic two-channel model. A 
bound state with energy $\nu$ is in the closed channel subspace $\mathcal{Q}$ (black curve). Atoms enter in the open channel subspace $\mathcal{P}$ (red curve) with incident energy $E$ and couple to the closed channel bound state.  By varying the applied magnetic field, the bound state can be tuned to resonance.}
\label{fig:PotentialsVacuum}
\end{figure}

 Considering an energetically open channel in the subspace $\mathcal{P}$ coupled to an energetically closed channel in the subspace $\mathcal{Q}$ with a bound state with energy $\nu$ as sketched in Fig.\;\ref{fig:PotentialsVacuum}, we split the coupled-channel Schr\"odinger equation into two components \cite{art:kokk}, such that
\begin{equation}
\label{eq:MatrixGeneral}
E\begin{bmatrix} \ket{\Psi_{\mathrm{P}}} \\ \ket{\Psi_{\mathrm{Q}}} \end{bmatrix} = \begin{bmatrix} \hat{H}_{\mathrm{PP}} & \hat{H}_{\mathrm{PQ}} \\ \hat{H}_{\mathrm{QP}} & \hat{H}_{\mathrm{QQ}}\end{bmatrix} \begin{bmatrix} \ket{\Psi_{\mathrm{P}}} \\ \ket{\Psi_{\mathrm{Q}}} \end{bmatrix},
\end{equation}
where we have employed the open- and closed-channel projection operators $\hat{P}$ and $\hat{Q}$ that project  the total scattering wave function $\Psi$ and the Hamiltonian $\hat{H} = \hat{H}^0 +\hat{V}$ onto the open- and closed-channel subspace respectively, such that $\ket{\Psi_{\mathrm{P}}} \equiv \hat{P}\ket{\Psi}$,$\ket{\Psi_{\mathrm{Q}}} \equiv \hat{Q}\ket{\Psi}$, $\hat{H}_{\mathrm{PQ}} \equiv \hat{P}\hat{H}\hat{Q}$ etc. 
\par 
Assuming that the energy $\nu$ of the closed channel bound state is close to the collision energy $E$ of the interacting particles in the subspace $\mathcal{P}$, we make the single resonance approximation and neglect the scattering states and other bound states in the closed channel subspace $\mathcal{Q}$.  Under this approximation, we can solve Eq.\;\eqref{eq:MatrixGeneral} for the closed-channel wave function $\ket{\Psi_{\mathrm{Q}}}$, finding 
\begin{equation}
\label{eq:ClosedIntermediate}
\ket{\Psi_{\mathrm{Q}}} = \frac{\ket{\phi}\bra{\phi}}{E-\nu} \hat{H}_{\mathrm{QP}}\ket{\Psi_{\mathrm{P}}},
\end{equation}
where $\ket{\phi}$ is the closed channel bare bound state. 
Substituting the previous relation into Eq.\;\eqref{eq:MatrixGeneral}, we eliminate $\ket{\Psi_{\mathrm{Q}}}$ from the expression for the open channel scattering wave function $\ket{\Psi_{\mathrm{P}}}$ such that
\begin{equation}
\label{eq:OpenIntermediate}
E\ket{\Psi_{\mathrm{P}}} = (\hat{H}^0_{\mathrm{PP}}+\hat{V}_{\mathrm{eff}})\ket{\Psi_{\mathrm{P}}},
\end{equation}
with 
\begin{equation}
\label{eq:EffectivePotential}
\hat{V}_{\mathrm{eff}} = \hat{V}_{\mathrm{PP}} +\hat{H}_{\mathrm{PQ}}\frac{\ket{\phi}\bra{\phi}}{E-\nu}\hat{H}_{\mathrm{QP}}.
\end{equation}
Equation \eqref{eq:OpenIntermediate} can be solved straightforwardly, yielding 
\begin{equation}
\label{eq:OpenIntermediate2}
\ket{\Psi_{\mathrm{P}}} = \ket{\psi_{\mathrm{P}}^+}+\frac{1}{E^+-\hat{H}_{\mathrm{PP}}} \frac{\hat{H}_{\mathrm{PQ}}\ket{\phi}\bra{\phi}\hat{H}_{\mathrm{QP}}}{E-\nu}\ket{\Psi_{\mathrm{P}}},
\end{equation}
where $\ket{\psi_{\mathrm{P}}^+}$ is the eigenstate of the direct open-channel interaction Hamiltonian $\hat{H}_{\mathrm{PP}}$ and $E^+ = E+i \delta$, with  $\delta \to 0^+$ to avoid singularities. \par
Throughout this work, we consider separable potentials
\begin{subequations}
\begin{align}
\hat{V}_{\mathrm{PP}} &= v\ket{\zeta}\bra{\zeta}, \label{eq:OpenChannelOp} \\
 \hat{V}_{\mathrm{PQ}} &= \beta \ket{\zeta}\bra{\zeta} \label{eq:ClosedChannelOp},
 \end{align}
\end{subequations} 
with form factors $\ket{\zeta}$ and open-channel and coupling potential strengths $v$ and $\beta$ respectively. Analogous to Ref.\;\cite{art:kokk}, we define the closed channel amplitude $\Phi_{Q
} \equiv \braket{\phi|\Psi_{\mathrm{Q}}}/\sqrt{2}$ for the system to be in the bound state. Using Eqs.\;\eqref{eq:ClosedIntermediate}-\eqref{eq:ClosedChannelOp}, Eq.\;\eqref{eq:MatrixGeneral} takes the form
\begin{subequations}
\begin{align}
\label{eq:OpenChannelEigenvalue}
 E  \Psi_{\mathrm{P}}(\mathbf{k}) &= \frac{\hbar^2 k^2}{m}\Psi_{\mathrm{P}}(\mathbf{k})+v \zeta(2\mathbf{k})\sum_{\mathbf{q}} \zeta^*(2\mathbf{q}) \Psi_{\mathrm{P}}(\mathbf{q})  \notag \\
& \quad + g\zeta(2\mathbf{k})\Phi_{\mathrm{Q}},
 \\
 \vspace{0.1cm}
\label{eq:ClosedChannelEigenvalue}
E  \Phi_{\mathrm{Q}}  &= \nu \Phi_{\mathrm{Q}}  +\frac{g}{2}\sum_{\mathbf{k}} \Psi_{\mathrm{P}} (\mathbf{k}) \zeta^*(2\mathbf{k}),
\end{align}
\end{subequations}
with potential interaction strength $g = \sqrt{2}\beta \braket{\phi|\zeta}$. Following Refs. \cite{art:MusolinoPairFormation,art:colussimk,PhysRevA.102.063314,musolino2021boseeinstein}, we choose a step-function form factor $\braket{\mathbf{k}/2|\zeta}  = \zeta(\mathbf{k}) = \Theta(\Lambda-\abs{\mathbf{k}}/2)$, with relative two-body momentum $\mathbf{k}$ and momentum cut-off $\Lambda$. The value of the cut-off is calibrated by matching the dimer binding energy of the two-channel model to the full coupled-channel dimer binding energy \cite{thomas} \footnote{For instance, in the case of the $B_0 = 155.04$ G resonance in $^{85}\text{Rb}$ \cite{Rb85Input}, we find $\Lambda = 0.6/r_{\mathrm{vdW}}$, with $r_{\mathrm{vdW}}$  the van der Waals length of $ ^{85}\text{Rb}$  \cite{art:chin}. We obtained the momentum cut-off $\Lambda$ by comparing the coupled-channels data from \cite{thomas} to a polynomial where we keep the zeroth order term in the low-energy expansion of $a(B)^{-1}$. Other resonances in different atomic species result in different calibrated values of the cut-off. For example, the calibration of the $B_0 = 402$ G resonance in $^{39}$K yields $\Lambda = 0.46/r_{\mathrm{vdW}}$. The calibrated value depends on the width of the considered resonance and approaches the value $\Lambda = 2\pi/\bar{a}$ \cite{art:colussimk,art:MusolinoPairFormation} with mean-scattering length $\bar{a} = 0.955 r_{vdw}$ in the single channel limit.}. The potential strengths $v$ and $g$ can be modelled using the following set of renormalization equations \cite{art:servaas_ram,PhysRevA.85.033616}
\begin{equation}
\label{eq:vpotential}
v = v_0 \Gamma = \frac{4 \pi \hbar^2 a_{\mathrm{bg}} }{m}\Gamma,
\end{equation} 
\vspace{-0.5cm}
\begin{equation}
 \label{eq:wpotential}
 g = g_0 \Gamma = \frac{(2\pi)^{3/2} \hbar^2}{\sqrt{R^*} \pi m} \Gamma,
 \end{equation}
with $\Gamma = \left(1-2\Lambda a_{\mathrm{bg}}/\pi\right)^{-1}$ and with renormalized interaction parameters $v$ and $g$. In order to investigate the effect of the separable potential interaction on the open channel wave function $\ket{\Psi_\mathrm{P}}$, we multiply Eq.\;\eqref{eq:OpenIntermediate2} from the left by $\bra{\mathbf{k}}\hat{V}_{\mathrm{eff}}$, where $\ket{\mathbf{k}}$ is an unscattered state. 
Exploiting the relation between the transition operator $\hat{T}$ and the scattering potential operator $\hat{T}\ket{\mathbf{k}} = \hat{V}_{\mathrm{eff}}\ket{\Psi_{\mathrm{P}}}$, we can find the following expression for the coupled channels two-body transition matrix in the case of a separable potential interaction
\begin{equation}
\label{eq:Tvac}
T(E) = v\zeta(2\mathbf{k}) \braket{\zeta|\psi_{\mathrm{P}}^+}+ \frac{\frac{g^2}{2} \abs{\braket{\zeta|\psi_{\mathrm{P}}^+}}^2}{E-\nu-\frac{g^2}{2}\braket{\zeta|\hat{G}_{\mathrm{P}}(E)|\zeta}},
\end{equation}
where we have introduced the open channel Green's operator $\hat{G}_{\mathrm{P}}(E) = (E-\hat{H}_0-v\ket{\zeta}\bra{\zeta})^{-1}$. 
We can use Eq.\;\eqref{eq:Tvac} in order to extract important scattering parameters and obtain the bound state energies from its poles, which we do presently. 
 \subsubsection{\label{subsec:2Bresonancewidthclassification} Vacuum classification of the resonance width}
As mentioned in Sec.\;\ref{sec:level1}, the description of narrower resonances, which in the vacuum classification corresponds to the limit $\abs{R^*/a_{\mathrm{bg}}} \gg 1$, requires the inclusion of finite range scales. These finite range scales affect the universal scaling of the system, which is clear from the effective range expansion \cite{taylor}
\begin{align}
\label{eq:EffectiveRangeExpansion}
k \cot[\delta_0(k)] \approx \frac{-1}{a}+\frac{1}{2}k^2 R_{\mathrm{eff}} + \mathcal{O}(k^4),
\end{align} 
where $\delta_0(k)$ is the s-wave phase shift which depends on the scattering length $a$ and the effective range $R_{\mathrm{eff}}$. 
The necessity to consider the energy-dependent correction to the phase shift follows directly from the inspection of the Breit-Wigner form of the phase shift for a non-resonant open-channel interaction, which can be expressed as \cite{PhysRevLett.108.250401} 
\begin{equation}
\label{eq:PhaseShiftGeneral}
\text{tan} [\delta_0(k)] = -k a_{\mathrm{bg}}-\frac{\Delta E}{E- \nu_0},
\end{equation} 
where $\Delta E = \hbar^2/m a_{\mathrm{bg}} R^*$ is the energy width on the vacuum length scale $a_{\mathrm{bg}}$ and where $\nu_0 = \delta\mu(B-B_0)$ is the energy detuning from the Feshbach resonance. Analogous to Ref.\;\cite{art:kokk}, the detuning $\nu_0$ can be related to the closed channel bound state energy $\nu$ as \cite{art:servaas_ram,PhysRevA.85.033616}
\begin{align}
\label{eq:detuning}
\nu = \nu_0 + \frac{m\Lambda}{4\pi^2 \hbar^2} g g_0.
\end{align}
Multiplying the numerator and denominator of the second term on the right-hand side of Eq.\;\eqref{eq:PhaseShiftGeneral} by a factor $m R^* /\hbar^2$, we can recognize that the term that scales with the square of the momentum and sets the energy scaling depends on the size of the resonance strength parameter $R^*$ \footnote{By applying Eq.\;\eqref{eq:DeltaB}, we find that Eq.\;\eqref{eq:PhaseShiftGeneral} can be rewritten as $\text{tan} [\delta_0(k)] = -k a_{\mathrm{bg}}-\left(k R^*- \frac{B-B_0}{\Delta B}\right)^{-1}$.}.
\par
In order to find the exact relation that connects the effective range to the resonance strength parameter $R^*$, we use the transition matrix as presented in Eq.\;\eqref{eq:Tvac} and relate it to the phase shift $\delta_0(E)$ as 
\begin{equation}
\label{eq:cotandT}
k \text{cot}[\delta_0(E)] = ik -\frac{4\pi \hbar^2}{m T(E)}. 
\end{equation}
In the case of step function separable potential interactions the expansion of Eq.\;\eqref{eq:cotandT} around $k\rightarrow 0$ and its comparison to Eq.\;\eqref{eq:EffectiveRangeExpansion} yield the following expression for the effective range 
\begin{equation}
\label{eq:effectiverangevacuum}
R_{\mathrm{eff}} = -2 R^* + \frac{4}{\pi \Lambda}+\frac{1}{a}\left(4a_{\mathrm{bg}} R^*-\frac{2 a_{\mathrm{bg}}^2 R^*}{a}\right),
\end{equation}
which reduces to $R_{\mathrm{eff}} \approx -2 R^*$ at unitarity and for large values of the momentum-space cut-off \cite{book:chapterservaas}.
 
 \subsubsection{\label{subsec:2BZparameter} Dimer wave function}
In the two-channel model, the bound state is dressed by the channel coupling. The closed-channel amplitude of this dressed bound state is quantified by the dimer wave function normalization factor, or $Z$-parameter. This parameter can be introduced consistently with Refs. \cite{rew:duinestoof,book:cohen} as 
\begin{equation}
\label{eq:CoupledChannelsMatrixVacuum}
\begin{bmatrix} \ket{\Psi_{\mathrm{P}}} \\ \ket{\Psi_{\mathrm{Q}} } \end{bmatrix} = \sqrt{Z} \begin{bmatrix} \hat{G}_{\mathrm{P}}(E_{\mathrm{D}})\beta \ket{\zeta} \braket{\zeta|\phi} \\ \ket{\phi} \end{bmatrix},
\end{equation} 
such that $\braket{\Psi|\Psi} = 1$ and where $E_{\mathrm{D}}$ is the bound state energy of the dressed dimer state.
In the case of the separable potential, the $Z$ parameter takes the form 
\begin{equation}
\label{eq:Zvacuum}
Z= \left[1+\frac{\frac{g^2}{2} \sum_{\mathbf{k}} \frac{\zeta(2\mathbf{k})}{(E_{\mathrm{D}}-\frac{\hbar^2 k^2}{m})^2} }{\left(1-v \sum_{\mathbf{k}} \frac{\zeta(2\mathbf{k})}{E_{\mathrm{D}}-\frac{\hbar^2 k^2}{m}}\right)^2}\right]^{-1}. 
\end{equation}
The inspection of the previous equation reveals that at unitarity, where $E_{\mathrm{D}} = 0$, $Z = 0$ \cite{art:falcostoff,PhysRevLett.95.020404}. Departing from unitarity, the $Z$-parameter is bounded to a maximum value of one, which corresponds to the dimer fully in the closed channel subspace. In Sec.\;\ref{subsec:embeddedZparameter}, the definition of the $Z$-parameter will be extended to include many-body effects of the background gas.
\subsection{\label{subsec:manybodyeq} The many-body problem}
We now proceed to model a uniform gas of identical bosons interacting via the separable cut-off potential interactions as introduced in Sec.\;\ref{subsec:CoupledChannelsTmatrix}. Using second quantization, the two-channel many-body Hamiltonian corresponds to
\begin{equation}
\begin{aligned}
\hat{H} = &\sum_{\mathbf{k}} \frac{\hbar^2 k^2}{2m}\hat{a}^{\dagger}_{\mathbf{k}} \hat{a}_{\mathbf{k}}+\sum_{\mathbf{k}}\left(\frac{\hbar^2 k^2}{4m}+\nu\right) \hat{b}^{\dagger}_{\mathbf{k}}\hat{b}_{\mathbf{k}}  \\ 
&+\frac{v}{2} \sum_{\mathbf{k},\mathbf{k'},\mathbf{q}} \zeta\left(\mathbf{k}-\mathbf{k'}+2\mathbf{q}\right)\zeta^*\left(\mathbf{k}-\mathbf{k'}\right)\hat{a}^{\dagger}_{\mathbf{k}+\mathbf{q}}\hat{a}^{\dagger}_{\mathbf{k'}-\mathbf{q}}\hat{a}_{\mathbf{k}'}\hat{a}_{\mathbf{k}} \\
&+ \frac{g}{2} \sum_{\mathbf{k},\mathbf{q}} \left[\zeta^*(\mathbf{2k})\hat{b}^{\dagger}_{2\mathbf{q}} \hat{a}_{\mathbf{q}+\mathbf{k}}\hat{a}_{\mathbf{q}-\mathbf{k}}+\text{H.c}\right],
\end{aligned}
\end{equation}
where we have neglected the interactions among closed-channel molecules \cite{PhysRevLett.83.2691,PhysRevLett.83.1550,PhysRevLett.86.1915,PhysRevA.63.031601,art:kokk}. We have introduced the open channel atomic operators $\hat{a}_{\mathbf{k}}$ and the closed channel molecular operators $\hat{b}_{\mathbf{k}}$, which can be expressed as  \cite{art:MusolinoPairFormation,art:altman}
\begin{align}
\label{eq:CompoundBosonicOperator}
\hat{b}^{\dagger}_{\mathbf{q}} = \sum_{\mathbf{k}} \frac{\phi(k)}{\sqrt{2}} \hat{c}^{\dagger}_{-\mathbf{k}+\mathbf{q}/2}\hat{c}^{\dagger}_{\mathbf{k}+\mathbf{q}/2},
\end{align}
where $\phi(k)$ is the wave number projection of the closed channel bound state $\ket{\phi}$ as introduced in Eq.\;\eqref{eq:ClosedIntermediate} and where $\hat{c}^{\dagger}_{\mathbf{k}}$ is a closed-channel atomic operator. We note that we are working in the limit where $\phi(k)$ is localized with respect to the density scales, such that $\hat{b}^{\dagger}_{\mathbf{q}}$ is a true bosonic molecular operator, contrary to the composite pair operator introduced in Ref.\;\cite{musolino2021boseeinstein}. \par 
We apply the Bogoliubov decomposition to the operators such that $\braket{a_0} = \psi_a$ ($\braket{b_0} = \psi_m$) and $\braket{a_{\mathbf{k} \neq 0} = 0}$ ($\braket{b_{\mathbf{k} \neq 0} = 0}$). Here $\psi_a$ ($\psi_m$) is the atomic (molecular) wave function. We neglect excitations of the molecular condensate \cite{PhysRevLett.83.2691,PhysRevLett.83.1550,PhysRevLett.86.1915,PhysRevA.63.031601,art:kokk}. 
\par  To model correlations in the many-body system, we perform a cumulant expansion \cite{art:kira,art:colussimk,art:MusolinoPairFormation}. This expansion separates clusters of correlated atoms and molecules within the many-body model. We truncate the cumulant expansion at the second order, such that we obtain the doublet model where only single- and two-particle correlations are considered. These correlations (or clusters) consist of the singlets $\psi_a$ and $\psi_m$, which represent the atomic and molecular condensate respectively, and the doublets $\rho_{\mathbf{k}} \equiv \braket{\hat{a}^{\dagger}_{\mathbf{k}}\hat{a}_{\mathbf{k}}}$ and $\kappa_{\mathbf{k}} \equiv \braket{\hat{a}_{-\mathbf{k}}\hat{a}_{\mathbf{k}}}$, which are the one-body density and pairing matrices for $k \neq 0$ respectively \cite{book:blaizot}. \par 
We limit the evolution of the initially pure atomic condensate to a time up to $t = 2 t_{\mathrm{n}}$. At longer times $\rho_{\mathbf{k}}$ begins to exceed unity and the exclusion of strongly-driven higher-order cumulants can no longer be justified \cite{PhysRevA.102.063314,art:colussimk,art:MusolinoPairFormation}. 
\par
Applying the cumulant model up to the doublet level and 
implementing the Heisenberg equation of motion $i \hbar  \, d \mathcal{\hat{O}}/dt = [\hat{\mathcal{O}},\hat{H}]$, we obtain the two-channel Hartree-Fock Bogoliubov (HFB) equations of motion \cite{book:blaizot}
\begin{eqnarray}
 i\hbar \dot{\psi}_a&=& v \left(|\zeta(0)|^2|\psi_a|^2+ 2\sum_{\kg \neq 0} |\zeta(\kg)|^2\rho_\kg\right) \psi_a \notag\\ 
 &+& v \psi_a^\ast \sum_{\kg\neq 0}\zeta(0)\zeta^\ast(2\kg) \kappa_\kg+g\zeta(0)\psi_m \psi_a^*,
 \label{eq:psia3}\\
 i\hbar \dot{\psi}_m &=& \nu \psi_{m} +\frac{g}{2}\psi_a^2 \zeta^*(0)+\frac{g}{2}\sum_{\kg\neq 0} \zeta^*(2\kg) \kappa_{\kg}, \label{eq:psim} \\
  \hbar \dot{\rho}_\kg &=& 2\mbox{Im}\left[\Delta_\kg\kappa_\kg^\ast\right],
  \label{eq:gn3}\\
    i\hbar\dot{\kappa}_\kg &=& 2h_\kg\kappa_\kg+ \left(1 + 2\rho_\kg\right)\Delta_\kg,\label{eq:kappa}
\end{eqnarray}
where
\begin{equation}
\label{eq:HFhamiltonian}
h_\kg= \frac{\hbar^2 k^2}{2m} + 2v \left(|\zeta(\kg)|^2|\psi_a|^2 +\sum_{\qg\neq 0}|\zeta(\kg -\qg)|^2\rho_\qg\right),
\end{equation}
and 
\begin{equation}
 \Delta_\kg=v\zeta(2\kg)\left(\zeta^\ast(0)\psi_a^2 +\sum_{\qg\neq 0}\zeta^\ast(2\qg) \kappa_\qg\right)+g\zeta(2\kg) \psi_m,
\end{equation}
are the Hartree-Fock Hamiltonian and the pairing field, respectively. 
The HFB equations of motion conserve the total number of atoms $N=N_{op}+N_{cl}$, where $N_{op}$ consists of $N_a = V\abs{\psi_a}^2$ condensate atoms, with system volume $V = N/n$, and $N_{exc} =  V\sum_{\mathbf{k}} \rho_{\mathbf{k}}$ excitations and where $N_{cl} $ consists of $2N_m$ atoms, with $N_m$ the number of closed channel molecules $N_m = V\abs{\psi_m}^2$. 
\subsubsection{\label{subsec:resonancewidthclassification} Many-body classification of the resonance width}
As previously outlined in Sec.\;\ref{sec:level1}, the density can be used to derive a set of Fermi scales that quantify the properties of the unitary Bose gas. Exploiting the Fermi energy, we can then obtain the many-body analogue of the resonance width classification in terms of $k_{\mathrm{n}}R^*$ \cite{PhysRevLett.108.250401}. 
Similarly to the two-body classification scheme discussed in Sec.\;\ref{subsec:2Bresonancewidthclassification}, the energy width $\Delta E = \hbar^2 k_{\mathrm{n}}/m R^*$ on the Fermi
momentum scale $k_{\mathrm{n}}$ quantifies how fast the phase-shift varies as a function of the typical many-body energy. We observe that for $k_{\mathrm{n}} R^* \ll 1$, the phase shift varies slowly over a large range of energies, such that we can neglect the energy dependent term in Eq.\;\eqref{eq:EffectiveRangeExpansion}. This corresponds to the case of a broad resonance. Instead, for a rapid variation, where $k_{\mathrm{n}} R^* \gg 1$, we cannot neglect the energy dependent term in Eq.\;\eqref{eq:EffectiveRangeExpansion} and we quantify the resonance to be narrow \cite{PhysRevLett.108.250401}. 
In addition to examining the energy width, we can define the lifetime of the closed-channel molecules $\tau$ relative to the Fermi timescale $t_{\mathrm{n}}$ as 
\begin{align}
\label{eq:MolecularLifetime}
\frac{\tau}{t_{\mathrm{n}}} = \frac{E_{\mathrm{n}}}{\Delta E} = k_{\mathrm{n}} R^*,
\end{align}
where the linear scaling with $k_{\mathrm{n}} R^*$ implies longer lifetimes for narrower resonances in the many-body classification scheme. 
\par 
As pointed out in Ref.\;\cite{PhysRevLett.108.250401}, the vacuum and many-body classifications of the resonance width are not equivalent for all values of $a_{\mathrm{bg}}$ and $k_{\mathrm{n}}$. 
We can still satisfy the vacuum classification $\abs{R^*/a_{\mathrm{bg}}} \ll 1$ of a broad-resonance as introduced in Sec.\;\ref{subsec:2Bresonancewidthclassification} whilst observing the narrow-resonance behavior quantified by $k_{\mathrm{n}} R^* \gg 1$ for sufficiently large gas densities. 
\subsubsection{\label{subsec:PairCondensation} Pair condensation}
Extending the analysis performed in Ref.\;\cite{musolino2021boseeinstein} to the two-channel model, the closed-channel molecular fraction $\psi_m$ and the presence of the non-zero pairing matrix $\kappa_{\mathbf{k}}$ introduced in Sec.\;\ref{subsec:manybodyeq} signal off-diagonal long-range ordering (ODLRO) and pair condensation. Interestingly, these pairs could be investigated experimentally by using a rapid quenching procedure towards the weakly interacting regime along the lines of Refs. \cite{art:hodby,art:MusolinoPairFormation}, which maps the condensed pairs onto true molecules.
Isolating the atomic condensate from the fluctuations in order to omit the ODLRO that arises trivially due to the presence of the atomic condensate \cite{KIRA2014200,ODLROYang}, we define the following two-body density matrix as
\begin{align}
\label{eq:rho2Def}
\underline{\underline{\rho}}^{(2)}(\mathbf{r}_1',\mathbf{r}_2',\mathbf{r}_1,\mathbf{r}_2) = \braket{\delta \hat{\underline{\psi}}^{\dagger}(\mathbf{r}_1')\delta\hat{\underline{\psi}}^{\dagger}(\mathbf{r}_2')\delta\hat{\underline{\psi}}(\mathbf{r}_1)\delta\hat{\underline{\psi}}(\mathbf{r}_2)}, 
\end{align}
where $\delta \hat{\underline{\psi}}$ is the two-channel vector containing the open and closed channel fluctuations \footnote{We can generally define $\delta\hat{\psi}$ as $\begin{bmatrix} \psi_{P}-\psi_a ,& \psi_{Q}-\psi_c \end{bmatrix}^{\text{T}}$, with $\psi_c$ a closed channel atomic condensate. Next, by using the single resonance approximation and applying Eq.\;\eqref{eq:CompoundBosonicOperator} we retain only a small portion of the closed channel part and rewrite $\braket{\delta\hat{\psi}_Q\delta\hat{\psi}_Q(\mathbf{r})} = \sqrt{2}\phi(r)\braket{\hat{b}_0}$.}. Generalizing Refs. \cite{ODLROYang,musolino2021boseeinstein} we spectrally decompose Eq.\;\eqref{eq:rho2Def}, such that 
\begin{align}
\label{eq:rho2DefSD}
&\underline{\underline{\rho}}^{(2)}(\mathbf{r}_1',\mathbf{r}_2',\mathbf{r}_1,\mathbf{r}_2;t) = \notag \\ 
& \sum_{\nu} N^{(2)}_{\nu}(t) \underline{\varphi}_{\nu}^{(2)}(\mathbf{r}_1',\mathbf{r}_2',t) \underline{\varphi}_{\nu}^{(2)*}(\mathbf{r}_1,\mathbf{r}_2,t),
\end{align}
where $\underline{\varphi}_{\nu}^{(2)}$ represent the two-component orthonormal eigenvectors with  eigenvalues  $N^{(2)}_{\nu}$.
In the case of ODLRO, we expect $\underline{\underline{\rho}}^{(2)}(\mathbf{r}_1',\mathbf{r}_2',\mathbf{r}_1,\mathbf{r}_2;t)$ to be non-zero in the long-range limit where $\abs{\sum_{i=1}^2 \mathbf{r}_i-\mathbf{r}_i'}/2 \rightarrow \infty$. 
Applying the second order cumulant expansion as presented in Sec.\;\ref{subsec:manybodyeq} to the LR limit of Eq.\;\eqref{eq:rho2Def}, only the eigenstates with the anomalous contractions $\braket{\delta \hat{\underline{\psi}}^{\dagger}(\mathbf{r}_1')\delta\hat{\underline{\psi}}^{\dagger}(\mathbf{r}_2')} \braket{\delta\hat{\underline{\psi}}(\mathbf{r}_1)\delta\hat{\underline{\psi}}(\mathbf{r}_2)}$ remain, since the Hartree and the Fock terms, defined as $\braket{\delta \hat{\underline{\psi}}^{\dagger}(\mathbf{r}_1')\delta\hat{\underline{\psi}}(\mathbf{r}_1)} \braket{\delta\hat{\underline{\psi}}^{\dagger}(\mathbf{r}_2')\delta\hat{\underline{\psi}}(\mathbf{r}_2)}$ and $\braket{\delta \hat{\underline{\psi}}^{\dagger}(\mathbf{r}_1')\delta\hat{\underline{\psi}}(\mathbf{r}_2)} \braket{\delta\hat{\underline{\psi}}^{\dagger}(\mathbf{r}_2')\delta\hat{\underline{\psi}}(\mathbf{r}_1)}$ respectively, typically vanish for separations that exceed a few Fermi lengths \cite{book:Leggett}. Consequently, we find that
\begin{align}
\label{eq:rho2DefSD2}
&\underline{\underline{\rho}}^{(2)}(\mathbf{r}_1',\mathbf{r}_2',\mathbf{r}_1,\mathbf{r}_2;t) \underrel{\text{long range}}{=} \notag \\ 
&  N^{(2)}_{0}(t) \underline{\varphi}_0^{(2)}(\mathbf{r}_1',\mathbf{r}_2',t) \underline{\varphi}^{(2)*}_0(\mathbf{r}_1,\mathbf{r}_2,t),
\end{align}
with the associated pair wave function
\begin{align}
\label{eq:PairWaveFunctionr}
\underline{\varphi}_0^{(2)}(\mathbf{r},t) = \frac{1}{\sqrt{N_0^{(2)}(t)}} \begin{bmatrix} \sum_{\mathbf{k}}e^{i\mathbf{k} \cdot \mathbf{r}} \kappa_{\mathbf{k}} \\ \sqrt{2} \phi(\mathbf{r}) \psi_m \end{bmatrix},   
\end{align}
and the macroscopic eigenvalue 
\begin{align}
\label{eq:N02Eigenvalue}
N_0^{(2)}(t) = \sum_{\mathbf{k}} \abs{\kappa_{\mathbf{k}}}^2 + 2\abs{\psi_m}^2.
\end{align}
Contrary to the atomic condensate fraction that can be derived from the one-body density matrix, the eigenvalue $N_0^{(2)}$ cannot be directly related to the pair condensate fraction. This is a consequence of the background gas of excitations that Bose-enhance interactions and thereby violate the bosonic commutation relations as discussed in Ref. \cite{musolino2021boseeinstein}. As a result, we need to renormalize the eigenvalue $N_0^{(2)}$ in order to avoid over counting the pair condensate fraction. 
Following the renormalization procedure as presented in App.\;\ref{subsec:RenormalizationPairCondensate}, we find that the pair condensate fraction $n_0^{(2)}$ can be expressed as 
\begin{align}
\label{eq:PairCondensate}
n_0^{(2)}(t) = \frac{N_0^{(2)}(t)}{1+\left[2/N_0^{(2)}(t)\right] \sum_{\mathbf{k}\neq 0} \abs{\kappa_{\mathbf{k}}^2} \rho_{\mathbf{k}}}.
\end{align}	
We note that the renormalization factor in the denominator of Eq.\;\eqref{eq:PairCondensate} is identical to the renormalization factor presented in Ref.\;\cite{musolino2021boseeinstein}, where a similar analysis is performed for the single channel Bose gas. This is due to the presence of the background gas in the open channel. The evolution of the pair fraction for various values of $k_nR^*$ will be discussed in Sec.\;\ref{subsec:populationdistribution}. 
\subsubsection{\label{subsec:TwoBodyContactTheory} The dynamical two-body contact}
When two bosons in a quantum gas separated by a distance $r_{12} = \abs{\mathbf{r}_1-\mathbf{r}_2}$ approach each other, such that $\Lambda^{-1}\ll r_{12} \ll \{n^{-1/3},\abs{a},\lambda_{\mathrm{dB}},\text{etc.}\}$, the many-body wave function $\Psi_{\mathrm{MB}}(\mathbf{r}_1\sigma_1,\mathbf{r}_2\sigma_2,...,\mathbf{r}_N\sigma_N)$ can be factorized as \cite{art:Werner2009}
\begin{align}
\label{eq:ManyBodyCoupled}
\Psi_{\mathrm{MB}}(\mathbf{r}_1\sigma_1,..,\mathbf{r}_N\sigma_N) \approx \phi_{2B}(r_{12})\mathcal{A}(\mathbf{R}_{12}\sigma_P,..,\mathbf{r}_N\sigma_N),
\end{align}
with center of mass coordinate $\mathbf{R}_{12} = \frac{m_1\mathbf{r}_1+m_2\mathbf{r}_2}{m_1+m_2}$ and channel index $\sigma_i$. The above factorization is a starting point in the derivation of a set of universal relations related to the two-body, or Tan contact \cite{TAN20082952,TAN20082971,TAN20082987,PhysRevA.86.053633,PhysRevA.91.013616} $\mathcal{C}_2$, defined as \footnote{We have normalized our momentum distribution $\sum_{\mathbf{k}}\rho_{\mathbf{k}}$ using the particle density $n$ instead of the particle number $N$. As a result, we are analysing the contact density $\mathcal{C}_2$.}
\begin{equation}
\label{eq:ContactOriginal}
\mathcal{C}_2 \hspace{-1cm} \underrel{\scriptstyle\small \Lambda \gg k \gg \{k_{\mathrm{n}}, a^{-1},\lambda^{-1}_{\mathrm{dB}},\text{etc.}\}}{\equiv} \hspace{-1cm}  k^4 \rho_{\mathbf{k}}.
\end{equation} 
Since the two-body contact effectively measures the probability for pairs of atoms to be close together, it is a valuable parameter in the analysis of the unitary Bose gas.  
In our doublet model the asymptotic scaling behavior described by Eq.\;\eqref{eq:ContactOriginal} emerges for momenta $k$ in the universal regime compared to the system specific length scales $a$, $k_{\mathrm{n}}$ and $R_{\mathrm{eff}}$. 
The set of HFB equations presented in Sec.\;\ref{subsec:manybodyeq} then allow us study the time evolution of $\rho_{\mathbf{k}}$ and hence to compute the two-body contact. 
Alternatively, by integration over the regularized part of the wave function $\mathcal{A}(\mathbf{R}_{12}\sigma_P,..,\mathbf{r}_N\sigma_N)$ introduced in Eq.\;\eqref{eq:ManyBodyCoupled}, it is possible to express the contact in terms of the open-open channel component of the two-body correlation function $ g_{\mathrm{PP}}^{(2)}(\mathbf{r},\mathbf{r}') = \braket{\hat{\psi}_{\mathrm{P}}^{\dagger}(\mathbf{r})\hat{\psi}_{\mathrm{P}}^{\dagger}(\mathbf{r}')\hat{\psi}_{\mathrm{P}}(\mathbf{r})\hat{\psi}_{\mathrm{P}}(\mathbf{r}')}$, such that 
\begin{align}
\label{eq:CorrelationFunctioncoupled}
g_{\mathrm{PP}}^{(2)}(\mathbf{r},\mathbf{r}') \hspace{-1cm}& \underrel{\Lambda^{-1}\ll r_{12} \ll \{n^{-1/3},\abs{a},\lambda_{\mathrm{dB}},\text{etc.}\}}{\approx} \hspace{-1cm} \abs{\phi_{2B}(r\rightarrow 0)}^2 \frac{\mathcal{C}_2}{(4\pi)^2}.
\end{align}
Relating the two-body wave function to the open-channel wave function as $\phi_{2B}(r) = -a^{-1} \lim_{E\rightarrow 0} \Psi_P(r)$ \cite{WernerCastinI}, it is possible to compute the well-defined zero range ($\Lambda \rightarrow \infty)$ and zero energy limit of the two-body wave function \footnote{we would have to consider energy corrections in order to obtain the sub leading order contact}. 
Using $\hat{T}\ket{\mathbf{k}} = \hat{V}_{\mathrm{eff}}\ket{\Psi_P}$ with the effective potential as defined in Eq.\;\eqref{eq:EffectivePotential} we obtain
\begin{align}
\left(v-\frac{g^2}{2\nu}\right)\phi_{2B}(r\rightarrow 0) \underset{\stackrel{E\rightarrow 0}{ \Lambda \rightarrow \infty}}{\approx} -\frac{4\pi \hbar^2}{m},
\end{align} 
such that the two-body contact can be expressed as
\begin{align}
\label{eq:CorrelationGeneralized}
\mathcal{C}_2 = \frac{m^2}{\hbar^4}\braket{\left(v-\frac{g^2}{2\nu}\right)^2 \hat{\psi}_P^{\dagger} \hat{\psi}_P^{\dagger} \hat{\psi}_P \hat{\psi}_P (r \rightarrow 0)}.
\end{align}
As derived in App.\;\ref{subsec:GeneralizedTanContact}, Eq.\; \eqref{eq:CorrelationGeneralized} is the bosonic version of the \textit{generalized} Tan relation introduced in Ref.\;\cite{PhysRevA.78.053606}. 
Contrary to its single channel analogue, which is retrieved for $g=0$ \cite{PhysRevLett.100.205301} and only holds in the broad resonance limit, the expression for the generalized contact is valid for all considered values of $k_{\mathrm{n}}R^*$. In addition, as outlined in App.\;\ref{subsec:GeneralizedTanContact}, the zero-range limit of Eq.\;\eqref{eq:CorrelationGeneralized} can be directly related to the set of cumulants as presented in Sec.\;\ref{subsec:manybodyeq}. This facilitates the analysis of the contact as extracted from the tail of the momentum distribution as well as computed using the generalized Tan relation. The dynamics of this quantity are the subject of Sec.\;\ref{subsec:TwoBodyContact}.
\subsection{\label{subsec:Embeddedtwobodyinteractions} Embedded two-body  interactions}
Having discussed the two-body as well as the many-body problem in terms of a two channels model, we now proceed to study the resonance-width dependent physics of two-body interactions embedded in a many-body environment. This allows us to probe the effect of the medium on two-body interactions and relate these effects to the observation of non-zero values of the Z-parameter in the unitary Fermi gas \;\cite{art:stoof2009,PhysRevLett.95.020404}. 
\vspace{0.4cm}
\subsubsection{\label{subsec:Embeddedtwobodytransitionmatrix} The embedded two-body transition matrix}  
\vspace{-0.1cm}
Following Ref.\;\cite{art:colussimk}, we start our analysis of the embedded two-body interactions by extending the two-body transition matrix in vacuum to its embedded analogue. 
To this extend we decompose $\kappa_{\mathbf{k}}$ and $\psi_m$ in terms of a complete basis set with open- and closed channel wave functions $\Psi^R_{\mathrm{P},\mu}(\mathbf{k})$ and $\Phi_{\mathrm{Q},\mu} $ respectively, such that we find
\begin{equation}
\label{eq:BasisMatrix}
\begin{bmatrix}
\kappa_{\mathbf{k}} \\ \psi_m \end{bmatrix} 
= \sum_{\mu} c_{\mu}(t) \begin{bmatrix} \Psi^{R}_{\mathrm{P},\mu}(\mathbf{k}) \\ \Phi_{\mathrm{Q},\mu} \end{bmatrix}, 
\vspace{-0.15cm}
\end{equation}
with $c_{\mu}(t)= c_{\mu} e^{-i E_{\mu} t/\hbar}$ and  $\Phi_{\mathrm{Q},\mu} = \braket{\phi|\Psi_{\mathrm{Q},\mu}}/\sqrt{2}$, analogous to the definition of the amplitude $\Phi_{\mathrm{Q}}$ in Sec.\;\ref{subsec:CoupledChannelsTmatrix}.  
In this model we  treat the density effects as quasi-stationary \cite{art:kira,kira_koch_2011,art:colussimk}. The subscript $R$ introduced in the open channel wave function $\Psi_{\mathrm{P},\mu}^R(\mathbf{k})$ indicates the usage of right eigenvectors. The origin of the asymmetry will be discussed shortly. 
Using Eq.\;\eqref{eq:CompoundBosonicOperator}, we recognize  that $\psi_m$ can be interpreted as the closed channel analogue of the pairing matrix. The coefficients $c_{\mu}$ then tell us how much of the total pairing matrix, consisting of open- and closed channel contributions, is contained in a dimer-basis state $\mu$. 
\par 
Applying the quasi-stationary approximation and neglecting the source terms related to the presence of the atomic condensate, we use Eq.\;\eqref{eq:BasisMatrix} to obtain the following set of two channel eigenvalue equations 
\begin{widetext}
\begin{subequations}
\begin{align}
\label{eq:kappaQuasiStationary}
 E_{\mu} \Psi^{R}_{\mathrm{P},\mu}(\mathbf{k}) &= 2h_{\mathbf{k}} \Psi^{R}_{\mathrm{P},\mu}(\mathbf{k})+(1+2\rho_{\mathbf{k}})\zeta(2\mathbf{k}) \left(v \sum_{\mathbf{q}} \Psi^{R}_{\mathrm{P},\mu}(\mathbf{q})\zeta^*(2\mathbf{q}) + g \Phi_{\mathrm{Q},\mu}\right), \\
\label{eq:psiMQuasiStationary}
 E_{\mu} \Phi_{\mathrm{Q},\mu} &= \nu \Phi_{\mathrm{Q},\mu} +\frac{g}{2}\sum_{\mathbf{k}} \Psi^{R}_{\mathrm{P},\mu} (\mathbf{k}) \zeta^*(2\mathbf{k}).
 \end{align}
\end{subequations}
\end{widetext}
Equations \eqref{eq:kappaQuasiStationary} and \eqref{eq:psiMQuasiStationary} should be compared to the two-body eigenvalue equations as presented in Eqs.\;\eqref{eq:OpenChannelEigenvalue} and \eqref{eq:ClosedChannelEigenvalue}. Here the effective interaction potential operator $\hat{V}_{\mathrm{eff}}$ that has been introduced in Eq.\;\eqref{eq:EffectivePotential} is replaced by the interaction potential operator $\hat{\mathcal{V}}_{\mathrm{eff}}$, where $\mathcal{\hat{V}}_{\mathrm{eff}} = \hat{B} \hat{V}_{\mathrm{eff}}$, with $\hat{B}$ the Bose-enhancement operator $\bra{\mathbf{k},\mathbf{k'}}\hat{B}= (1+\rho_{\mathbf{k}}+\rho_{\mathbf{k'}})\bra{\mathbf{k},\mathbf{k'}}$ \cite{art:colussimk}.
The Bose-enhancement of open channel excitations causes the asymmetry of the open channel eigenvalue equation. \par
Additionally, the kinetic energy term in Eq.\;\eqref{eq:OpenChannelEigenvalue} is represented by the Hartree-Fock term $h(\mathbf{k}) \approx \hbar^2 k^2/2m +2v n_{op}$ in Eq.\;\eqref{eq:kappaQuasiStationary}, meaning that the energy in the open channel is mean-field shifted. Therefore, analogously to Ref.\;\cite{art:colussimk}, we quantify the binding energy and the detuning relative to the mean-field energy shifted threshold, such that $\mathcal{E}_{\mu} \equiv E_{\mu} - 4vn_{op}$ and  $\mathpzc{v} \equiv \nu -4vn_{op}$ respectively. 
\par
By using the effective potential interaction operator $\mathcal{\hat{V}}_{\mathrm{eff}}$, we can straightforwardly introduce the \textit{embedded} transition operator as $\mathcal{\hat{T}}\ket{\mathbf{k}} = \mathcal{\hat{V}}_{\mathrm{eff}}\ket{\Psi_{P,\mu}^{R}}$ and, for the separable potential introduction, obtain the embedded transition matrix 
\begin{widetext}
\begin{equation}
\label{eq:EmbeddedT}
\mathcal{T} = \braket{\mathbf{k}|\hat{B} v|\zeta}\braket{\zeta|\psi_{\mathrm{P},\mu}^{R,+}}+\frac{\frac{g^2}{2} \braket{\psi_{\mathrm{P},\mu}^{R,-}|\hat{B}|\zeta} \braket{\zeta|\psi_{\mathrm{P},\mu}^{R,+}}}{\mathcal{E}_{\mu}-\mathpzc{v}-\frac{g^2}{2}\braket{\zeta|\mathcal{\hat{G}}_{\mathrm{P}}(\mathcal{E}_{\mu}) \hat{B} |\zeta}},
\end{equation}
\end{widetext}
where $\mathcal{\hat{G}}_{\mathrm{P}}$ is the embedded open channel Green's operator $\mathcal{\hat{G}}_{\mathrm{P}}(\mathcal{E}_{\mathrm{\mu}}) = (\mathcal{E}_{\mathrm{\mu}}-\hat{B}v\ket{\zeta}\bra{\zeta})^{-1}$. Equation \eqref{eq:EmbeddedT} is related to the many-body transition operator $\hat{T}^{MB}$ \cite{Stoof:2009kfa} according to $\mathcal{\hat{T}} = \hat{B}\hat{T}^{MB}$. \par
One of the dimer-basis states $\mu = D$ corresponds to the embedded dimer. The energy of this dimer $\mathcal{E}_{\mathrm{D}}$ can be extracted from the pole of Eq.\;\eqref{eq:EmbeddedT}. Through the analysis of the zero energy limit of the embedded transition matrix, we can extract an embedded analogue of the scattering length $\mathpzc{a}$ and the effective range $\mathpzc{R_{\mathrm{eff}}}$, such that \footnote{Equation \eqref{eq:Texpansion} is valid for interactions with constant form factors.}
\begin{equation}
\label{eq:Texpansion}
\mathcal{T}_{E\rightarrow 0} \approx \frac{4 \pi \hbar^2 \mathpzc{a}}{m}\left(1-ik \mathpzc{a}+\frac{\mathpzc{a} \mathpzc{R_{\mathrm{eff}}}-2 \mathpzc{a}^2}{2}k^2 + \mathcal{O}(k^3)\right).
\end{equation}
The evolution of these quantities as a function of the resonance width (or equivalently $k_{\mathrm{n}}R^*$) will be the investigated in more detail in Sec.\;\ref{subsec:TheEmbeddedDimer}. 
 \subsubsection{\label{subsec:embeddedZparameter} The embedded dimer wave function normalization factor}
As mentioned in Sec.\;\ref{subsec:2BZparameter}, the Z-parameter is zero at unitarity for the two-body model. However, a finite value of the Z-parameter was previously predicted and observed in the unitary Fermi gas\;\cite{art:stoof2009,PhysRevLett.95.020404}. This motivates us to analyze the embedded version of the Z-parameter. This parameter quantifies the division of the embedded dimer with energy $\mathcal{E}_{\mathrm{D}}$ amongst the open- and closed channel subspaces.Analogous to Eq.\;\eqref{eq:Zvacuum}, we define the $\mathcal{Z}$-parameter as
\begin{equation}
\label{eq:CoupledChannelsMatrixEmbedded}
\begin{bmatrix} \ket{\Psi^R_{\mathrm{P,D}}} \\ \ket{\Psi_{\mathrm{Q,D}}} \end{bmatrix} = \sqrt{\mathcal{Z}} \begin{bmatrix} \mathcal{\hat{G}}_{\mathrm{P}}(E_{\mathrm{D}}) \hat{B} \beta \ket{\zeta}\braket{\zeta|\phi} \\ \ket{\phi} \end{bmatrix},
\end{equation} 
such that $\braket{\Psi_{\mathrm{D}}|\Psi_{\mathrm{D}}}=1$ and we find that 
\begin{equation}
\label{eq:Zembedded}
\mathcal{Z} = \left[1+\frac{\frac{g^2}{2} \sum_{\mathbf{k}} \zeta(2\mathbf{k}) \frac{(1+2\rho_{\mathbf{k}})}{(\mathcal{E}_{\mathrm{D}}-\frac{\hbar^2k^2}{m})^2} }{\left(1- v \sum_{\mathbf{k}} \zeta(2\mathbf{k}) \frac{(1+2\rho_{\mathbf{k}})}{\mathcal{E}_{\mathrm{D}}-\frac{\hbar^2k^2}{m}}\right)^2}\right]^{-1}. 
\end{equation}
As the $\mathcal{Z}$-parameter can be linked to the change in $\delta\mu$ \cite{PhysRevLett.91.240402,art:falcostoff} as well as to the observed atom-loss in molecular probe experiments as investigated in Ref.\;\cite{PhysRevLett.95.020404}. The relation between the $\mathcal{Z}$-parameter and experimental observables makes it a valuable quantity to connect the theory of embedded dimers to experiments.  As such, we will analyze its resonance width dependent evolution in Sec.\;\ref{subsec:Z-parameter}. 
\subsubsection{\label{subsec:embeddedFparameter} The relative dimer state occupation}
In addition to analyzing how the dimer is distributed amongst the two subspaces, we also wish to quantify the relative importance of this dimer state to the total pairing field. Therefore, we introduce the relative dimer state occupation $\mathcal{F}_{\mathrm{D}}$ \cite{art:kira}, defined as 
\begin{equation}
\label{eq:FparamDefinition}
\mathcal{F}_{\mathrm{D}} = \frac{\abs{c_{D}}^2}{\sum_{\mu}\abs{c_{\mu}}^2},
\end{equation}
in terms of the coefficients $c_{\mu}$ of the dimer basis in Eq.\;\eqref{eq:MatrixGeneral}. 
In order to compute the numerator of Eq.\;\eqref{eq:FparamDefinition}, we  multiply Eq.\;\eqref{eq:BasisMatrix} from the left side by $\begin{bmatrix} [\Psi_{\mathrm{P,D}}^L(\mathbf{k})]^*  & 2[\Phi_{\mathrm{Q,D}}]^*/\sqrt{V} \end{bmatrix}$, sum over all $\mathbf{k}$ and compute the square of the absolute value, such that we find 
\begin{equation}
\label{eq:NumeratorF}
\abs{c_{\mathrm{D}}}^2  = \abs{\sum_{\mathbf{k}} \kappa_{\mathbf{k}} \frac{[\Psi_{\mathrm{P,D}}^{R}(\mathbf{k})]^*}{1+2\rho_{\mathbf{k}}} + \sqrt{2 \mathcal{Z}}\psi_m}^2,
\end{equation}
where we have used the normalization condition $\sum_{\mathbf{k}} [\Psi_{P,\lambda}^L(\mathbf{k})]^* \Psi_{\mathrm{P},\mu}^R (\mathbf{k}) + 2[\Phi_{Q,\lambda}]^*\Phi_{\mathrm{Q},\mu} = \delta_{\lambda,\mu}$, the relation between the right- and left-eigenvectors of the open channel subspace $\Psi_{\mathrm{P},\mu}^R(\mathbf{k}) = \left(1+2\rho_{\mathbf{k}}\right)\Psi_{\mathrm{P},\mu}^L(\mathbf{k})$ and where we have applied Eq.\;\eqref{eq:CoupledChannelsMatrixEmbedded}.  \par
Applying a similar strategy, we next compute the denominator of Eq.\;\eqref{eq:FparamDefinition} by multiplying Eq.\;\eqref{eq:BasisMatrix} from the left side by $\begin{bmatrix} [c_{\lambda} \Psi_{P,\lambda}^L(\mathbf{k})]^*  & 2[c_{\lambda} \Psi_{Q,\lambda}]^* \end{bmatrix}$, summing over all values of $\mathbf{k}$ and computing the square of the absolute value. Once more we exploit the normalization condition and find 
\begin{equation}
\abs{c_{\mu}}^2 = \sum_{\mathbf{k}} \frac{\left[c_{\mu}\Psi_{\mathrm{P},\mu}(\mathbf{k})\right]^*}{1+2\rho_{\mathbf{k}}} \kappa_{\mathbf{k}} + 2\left[c_{\mu} \Psi_{\mathrm{Q},\mu}\right]^* \psi_m .
\end{equation}
Summing the previous expression over all basis-states $\mu$ and using Eq.\;\eqref{eq:BasisMatrix} in order to rewrite the dimer-basis states in terms of cumulants, we obtain the following expression for the denominator of $\mathcal{F}_{\mathrm{D}}$
\begin{equation}
\sum_{\mu} \abs{c_{\mu}}^2 = \sum_{\mathbf{k}} \frac{\abs{\kappa_{\mathbf{k}}}^2}{1+2\rho_{\mathbf{k}}} + 2\abs{\psi_m}^2.
\end{equation}
The expression for $\mathcal{F}_{\mathrm{D}}$ as derived here is normalized at every time step and can be applied to two channel systems. Consequently, we can analyze how the relative dimer state occupation evolves as a function of the time for various values of the resonance width, presenting our results in Sec.\;\ref{subsec:Relative dimer-state occupation}.
\section{ \label{sec:comparison}Results}
Having outlined our model, we now analyze the results for the dynamics of the quenched unitary Bose gas over a range of $k_\mathrm{n}R^*$. In this study, we follow the procedure as outlined by Refs. \cite{art:MusolinoPairFormation,art:colussimk}, starting with a non-interacting pure atomic condensate and perform an effectively sudden quench to unitarity \footnote{The analysis is not limited to a fixed atomic species or resonance and can be applied generally for varying values of $k_\mathrm{n}R^*$. In our simulations, we consider a condensate of $^{85}\text{Rb}$ atoms at an experimentally relevant density of $n = 4 \times 10^{12} \, \text{cm}^{-3}$, such that $t_{\mathrm{n}}= 69.84 \, \mu s$. We model the Feshbach resonance located at $B_0 = 155.04$ G with magnetic field width $\Delta B = 10.7$ G and background scattering length $a_{\mathrm{bg}} = -443 \, a_0$ \cite{Rb85Input} and quench the magnetic field to resonance in a time span of $\tau_{in}= 0.072 \, t_{\mathrm{n}}$, such that the quench is effectively sudden. We mark the completion of the quench as $t=0$ and start our analysis with this broad-resonance system, where $k_{\mathrm{n}} R^* \approx 10^{-3} $, in order to verify the consistency of the coupled-channels model with the single-channel model as presented in Ref.\;\cite{art:colussimk} and then gradually increase the value of $k_{\mathrm{n}}R^*$.}. Whereas we vary the value of $k_{\mathrm{n}} R^*$, we keep the value of $a_{\mathrm{bg}}$ fixed in order to satisfy the diluteness criterion $na^3_{\mathrm{bg}} \ll 1$.  This means that our results can be universally extended to different atomic species in the dilute regime with the same values of $k_{\mathrm{n}}R^*$.  
In the following sections, we express all our results in terms of Fermi-units unless mentioned otherwise and indicate the dimensionless resonance widths in terms of the many-body classification $k_{\mathrm{n}} R^*$. 
\subsection{\label{subsec:populationdistribution} Population fractions}
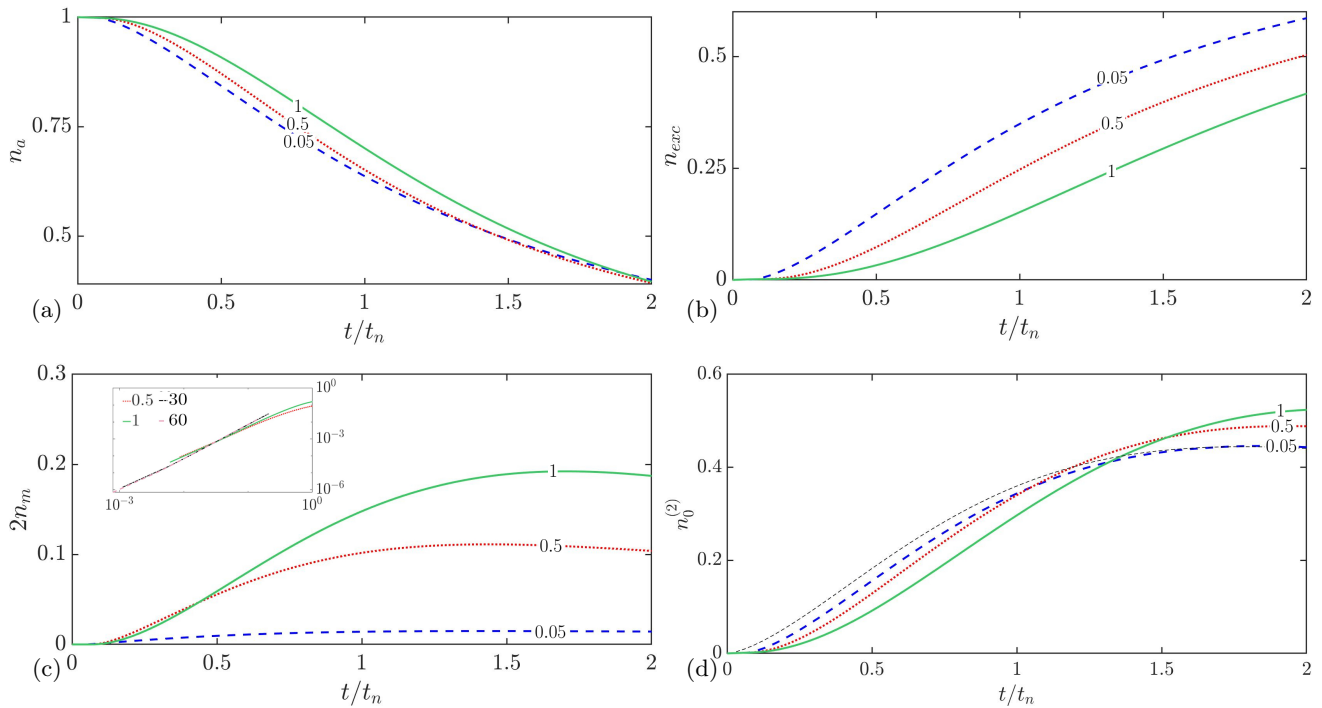
\begin{figure*}
      \begin{minipage}{.48\textwidth}
       \begin{tikzpicture}[>=latex]
\node at (0,0)
{\includegraphics[width=8.6cm]{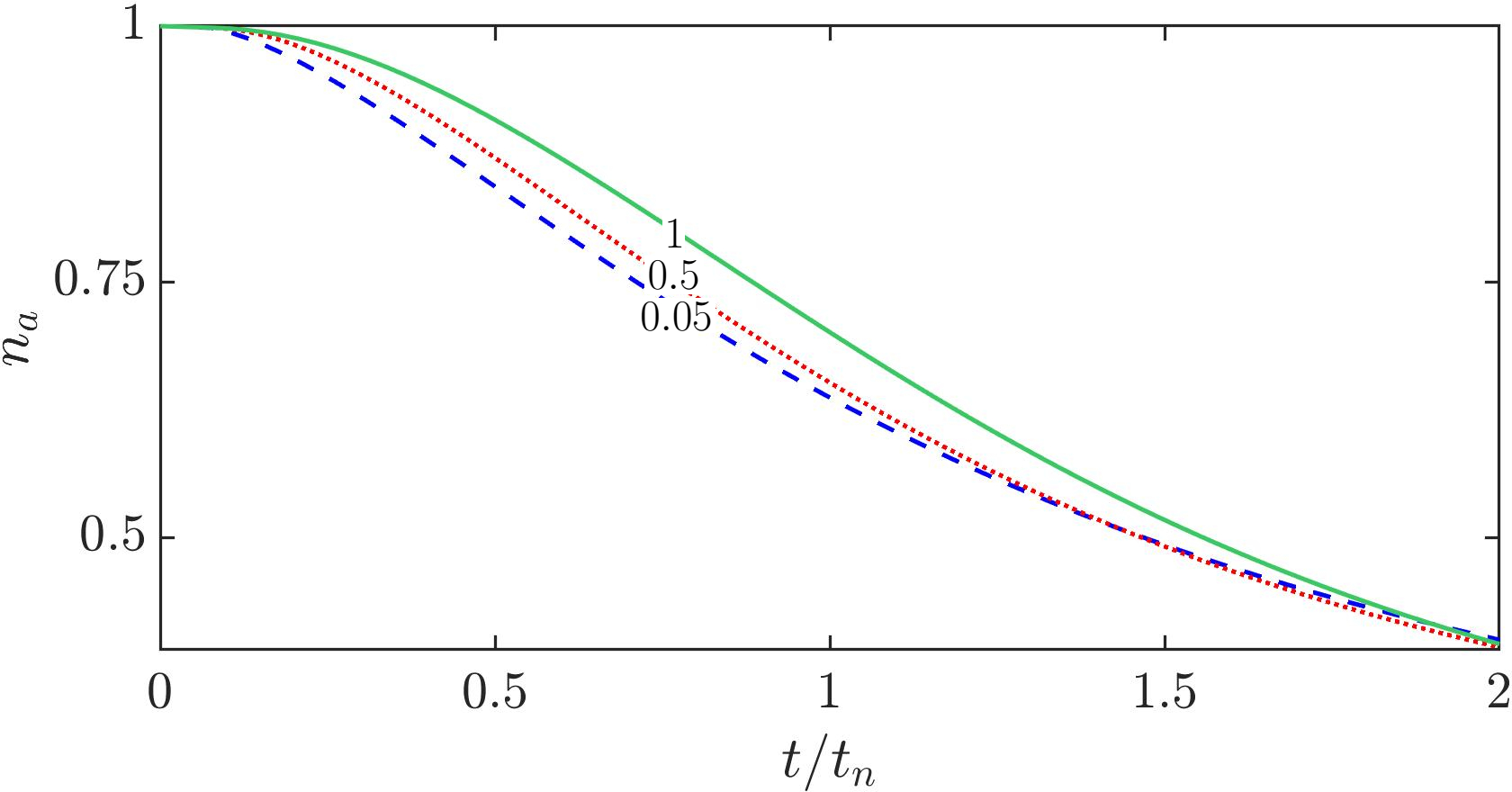}};
%\draw [->, thick] (0.5,1) -- (2.3,-0.1);
\node at (-3.8,-1.8) {(a)};
\end{tikzpicture}      
           \end{minipage}
      \begin{minipage}{.48\textwidth}
       \begin{tikzpicture}[>=latex]
\node at (0,0)
{\includegraphics[width=8.6cm]{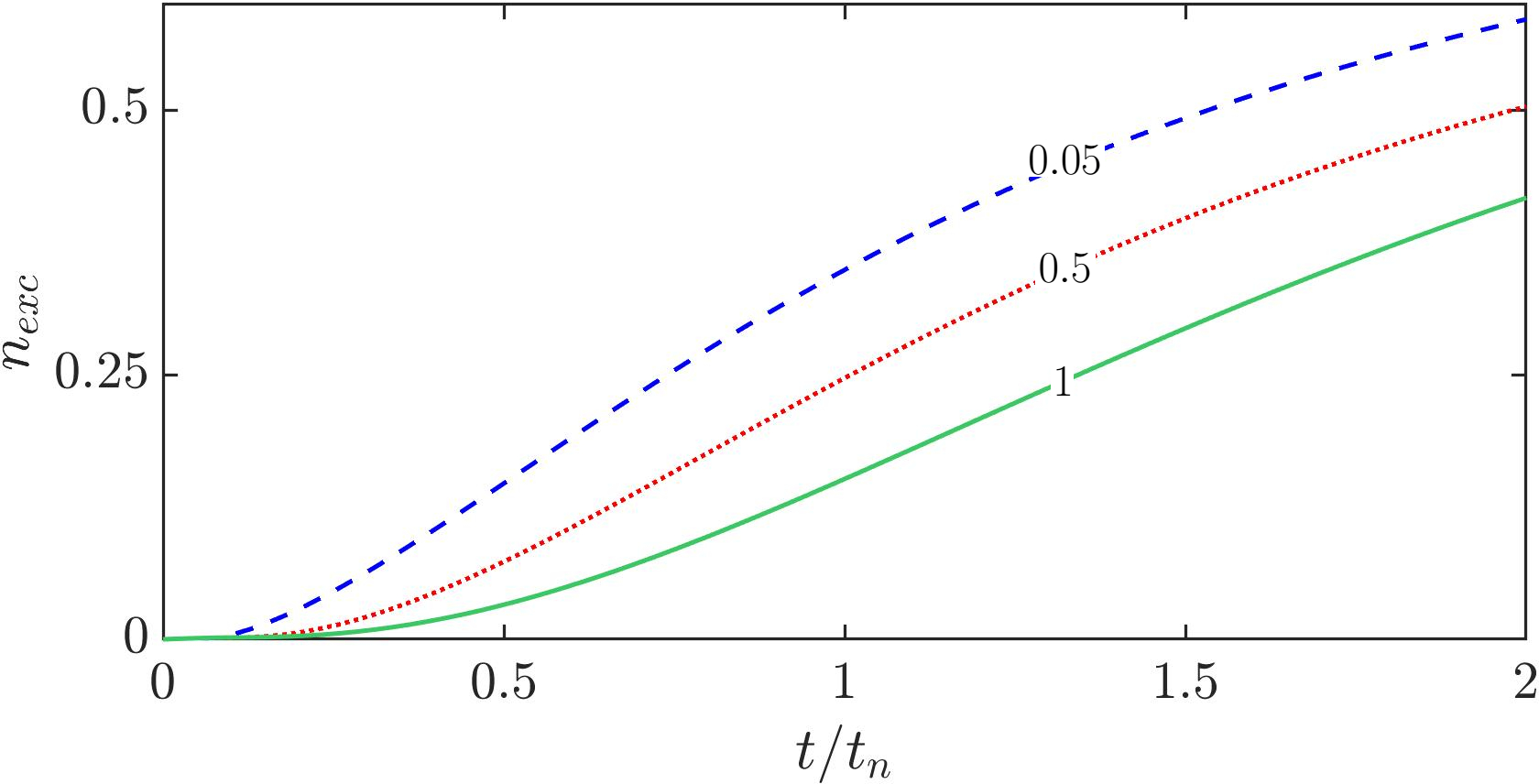}};
%\draw [->, thick] (0.5,1) -- (2.3,-0.1);
\node at (-3.8,-1.8) {(b)};
\end{tikzpicture}   
      \end{minipage}    
      \begin{minipage}{.48\textwidth}
       \begin{tikzpicture}[>=latex]
\node at (0,0)
{\includegraphics[width=8.6cm]{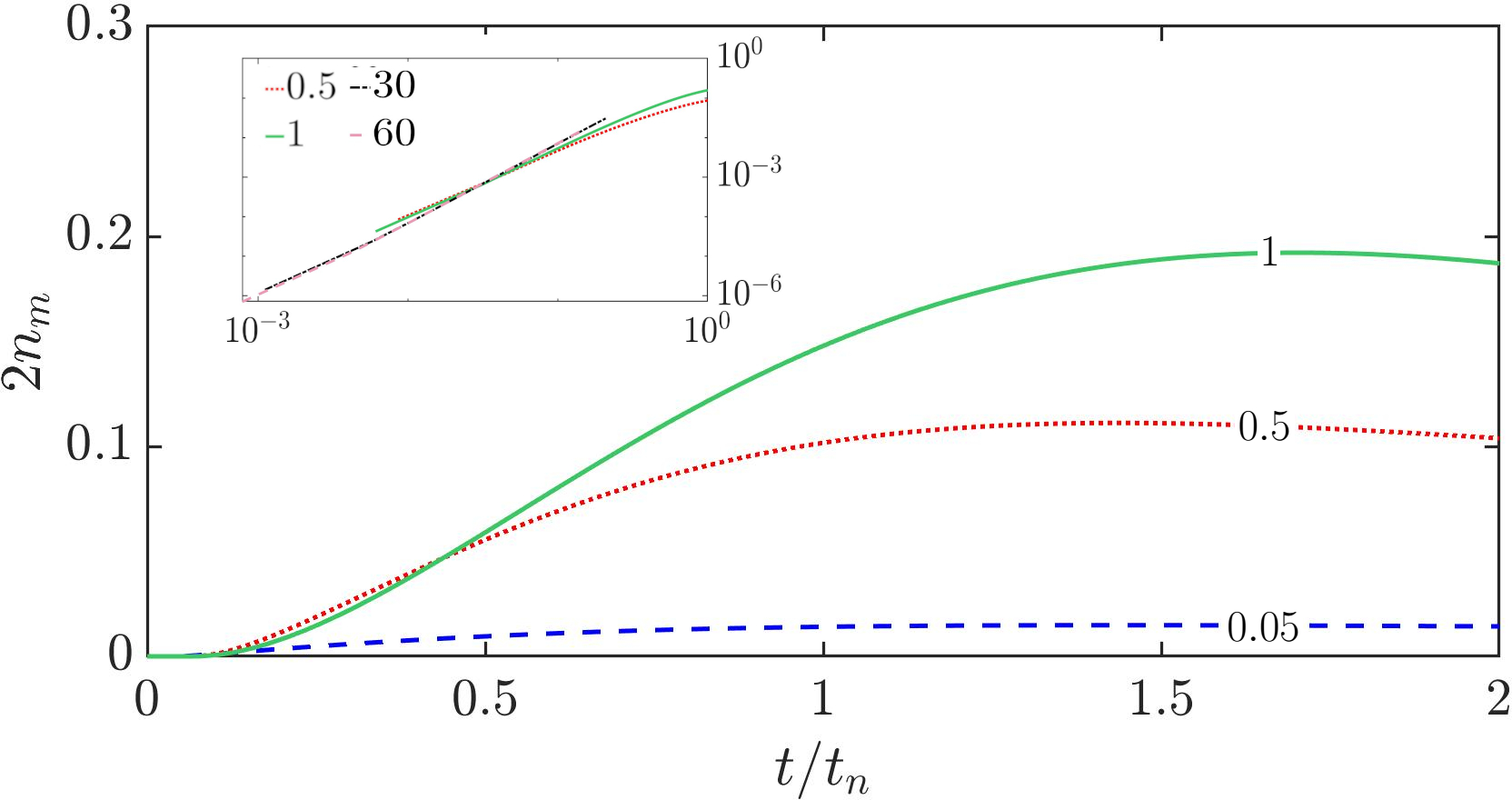}};
%\draw [->, thick] (0.5,1) -- (2.3,-0.1);
\node at (-3.8,-1.8) {(c)};
\end{tikzpicture} 
  \end{minipage}
      \begin{minipage}{.48\textwidth}
       \begin{tikzpicture}[>=latex]
\node at (0,0)
{\includegraphics[width=8.6cm]{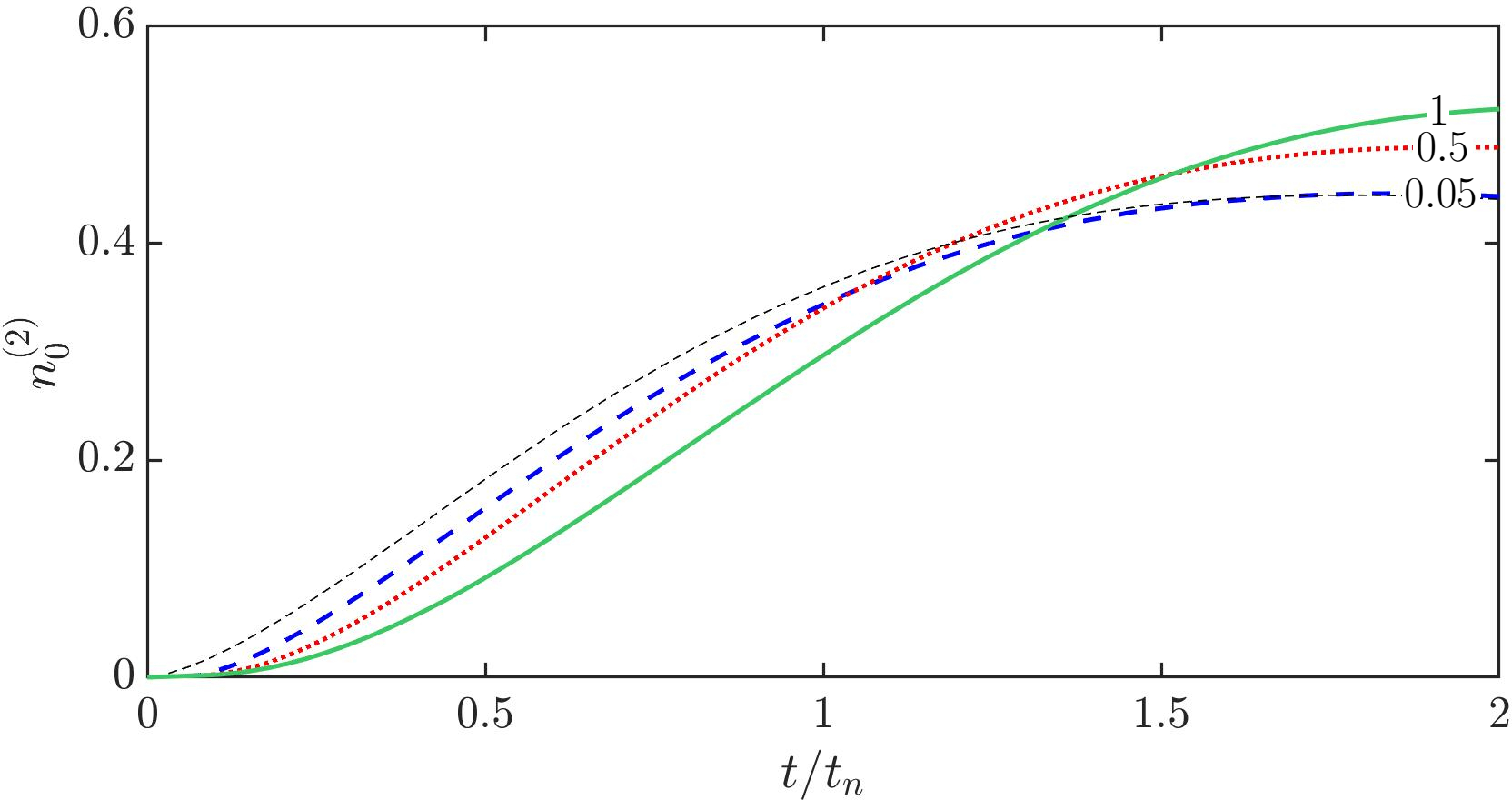}};
%\draw [->, thick] (0.5,1) -- (2.3,-0.1);
\node at (-3.8,-1.8) {(d)};
\end{tikzpicture}   
      \end{minipage}
             \caption{Population fractions at unitarity: (a) atomic condensate fraction $n_a$, (b) excited state fraction $n_{exc}$, (c) molecular condensate fraction $n_m$ and (d) pair condensate fraction $n_0^{(2)}$ as a function of the time $t/t_{\mathrm{n}}$ for various values of the dimensionless resonance width parameter $k_{\mathrm{n}} R^*$. The inset in (c) represents the molecular condensate fraction as a function of the time, with the time-axis rescaled with the embedded transition time $t_{*} = \sqrt{t_{\mathrm{n}}\tau}$. The legend in the inset indicates the resonance width in multiples of $k_{\mathrm{n}} R^*$. The dashed black line in (d) represents the single channel data as presented in Ref.\;\cite{musolino2021boseeinstein}. Our results are consistent with this calculation.} \label{fig:Population}
    \end{figure*}

Figure \;\ref{fig:Population} illustrates the (normalized) population dynamics as the time spent at unitarity progresses.
In particular Fig.\;\ref{fig:Population}(a) shows how the atomic condensate is quantum depleted. The atoms that leave the condensate can either form excitations (Fig.\;\ref{fig:Population}(b)) or closed channel molecules (Fig.\;\ref{fig:Population}(c)). 
Since the formed closed channel molecules have a shorter lifetime for broader resonances ($k_{\mathrm{n}} R^* \ll 1$), more excitations are formed at early times with respect to resonances with larger $k_{\mathrm{n}} R^*$. Additionally, the presence of background excitations Bose-enhances the production of extra excitations. Consequently, the excited state fraction as presented in Fig.\;\ref{fig:Population}(b) grows more rapidly for small $k_{\mathrm{n}} R^*$ at early times. \par 
Aiming to analyze the early-time dynamics of the molecular condensate on a more quantitative level, we consider the narrow-resonance ($k_{\mathrm{n}} R^* \gg 1$) and early-time limit ($t \ll t_{\mathrm{n}}$) of Eq.\;\eqref{eq:psim}. Since all atoms start out in the atomic condensate and Fig.\;\ref{fig:Population}(b) reveals that the excited state fraction remains limited at early times for resonances with large $k_{\mathrm{n}} R^*$, we approximate the atomic wave function as 
$\psi_a \approx \sqrt{n}$ and neglect the terms scaling with $\psi_m$ and $\kappa_{\mathbf{k}}$ in Eq.\;\eqref{eq:psim}. 
Under these approximations, we can integrate Eq.\;\eqref{eq:psim} with respect to the time and obtain
\begin{equation}
\label{eq:MolecularEarlyTimes}
2\abs{\psi_m}^2 \propto (t/t_{\ast})^2,
\end{equation}
where  $t_* = \sqrt{t_{\mathrm{n}} \tau}$ is the geometric mean of two relevant time scales. Here, $\tau = 2m R^*/k_{\mathrm{n}}\hbar$ is the lifetime associated to a molecule on the Fermi scale as discussed in Sec.\;\ref{subsec:CoupledChannelsTmatrix}. Physically $t_*$ represents the density-averaged time for open channel atoms to transition to closed channel molecules. Hence, we refer to $t_*$ as the mean transition time. The scaling of the early time dynamics of resonances in terms of the mean transition time $t_{*}$  is supported by the inset of Fig.\;\ref{fig:Population}(c), where the rescaling of the time-axis with $t_{*}$ results in the collapse of the molecular condensate fraction curves for sufficiently large values of $k_{\mathrm{n}} R^*$. 
\par Following the stage of rapid initial growth, Fig.\;\ref{fig:Population}(c) shows the saturation of the molecular condensate fraction at later times. 
The time it takes to reach the saturated value increases as a function of $k_{\mathrm{n}} R^*$. We interpret the delay in the observed saturation time to be a result of the increased mean transition time of closed channel molecules for narrower resonances. \par Once the molecular condensate fraction has saturated, the depletion of the atomic condensate fraction effectively solely results in the formation of new excitations. This process can be observed in Fig.\;\ref{fig:Population}(b), where we recognize that, at later times, the excited state fraction grows more rapidly for narrow resonances with respect to broader resonances. Furthermore, Fig.\;\ref{fig:Population}(a) shows how the difference between the atomic condensate fraction for broad and narrow resonances starts to decrease at later times. \par 
The dynamical evolution of the excited state fraction and the molecular condensate is echoed by the pair condensate fraction as presented in Fig.\;\ref{fig:Population}(d). We recognize that, for all considered resonance widths, a considerable fraction of the gas contributes to the pair condensate within the simulated time frame. In addition, this fraction is relatively insensitive to $k_{\mathrm{n}}R^*$ compared to the other fractions in Fig.\;\ref{fig:Population} and reflects the early time scaling law change observed in the molecular condensate fraction.
\subsection{\label{subsec:TwoBodyContact} Two-body contact}
By applying Eqs.\;\eqref{eq:ContactOriginal} and \eqref{eq:ContactGeneralizedFinal} to the doublet model we can compute the two-body contact, which encodes the probability of finding clustered particles at short distances, for different values of $k_{\mathrm{n}} R^*$.
\begin{figure}[t!]
\centering
\includegraphics[width=8.6cm]{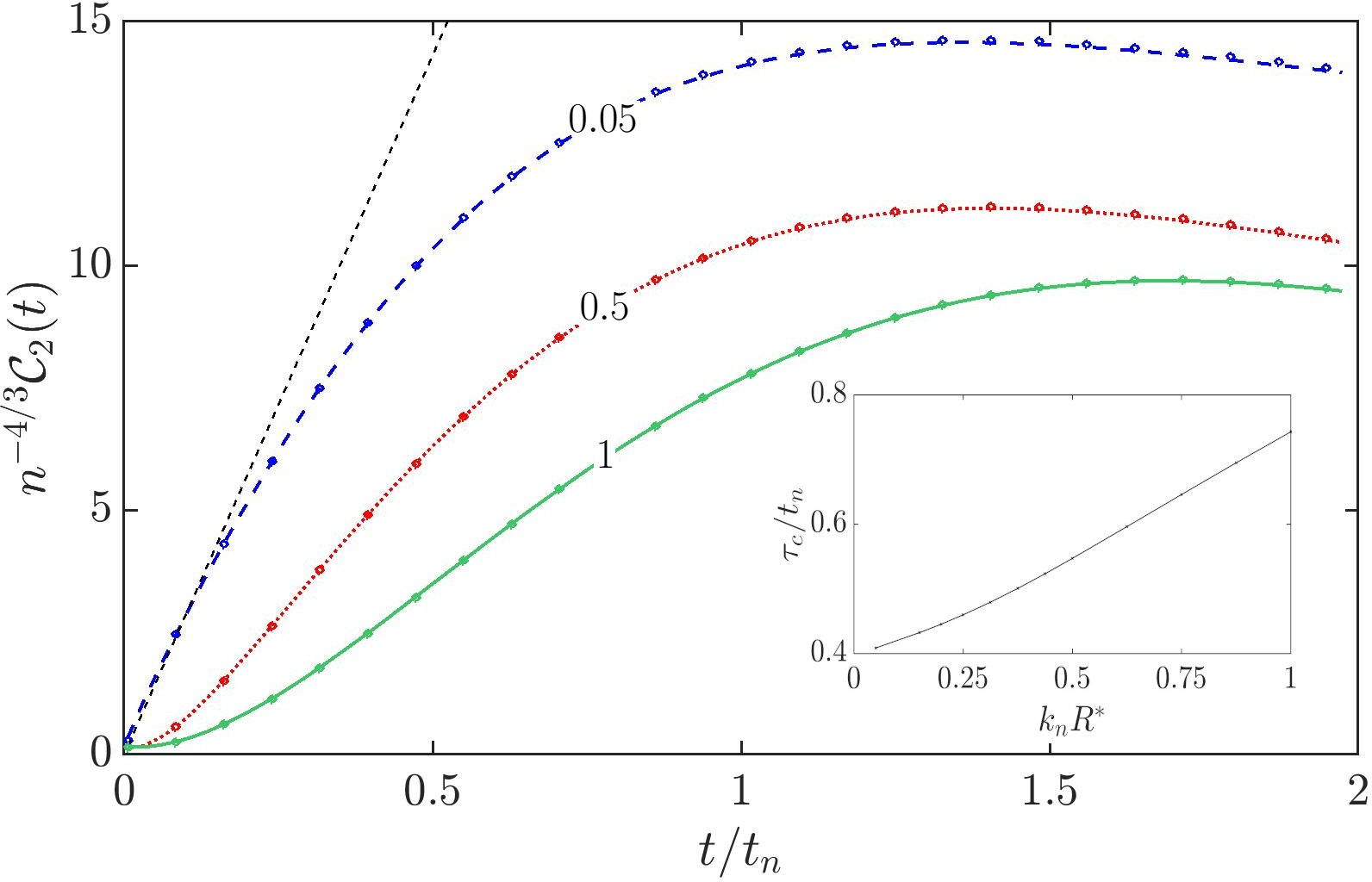}
\caption{Two-body contact rescaled with the density $n^{-4/3}$ fitted from the tail of the momentum distribution (full lines) and computed using the generalized Tan relation (points) as a function of the time for various values of the dimensionless resonance width $k_{\mathrm{n}}R^*$. The black line indicates the linear fit as presented in Eq.\;\eqref{eq:ContactLinear}. In the inset, the half-way time $\tau_c $ to reach the maximum value of the contact is plotted as a function of $k_{\mathrm{n}} R^*$.}
\label{fig:Contact}
\end{figure}

Figure \ref{fig:Contact} reveals that the contact obtained using the generalized Tan relation agrees with the contact obtained using the tail of the momentum distribution over the whole range of considered resonance widths. Furthermore, Fig.\;\ref{fig:Contact} shows that the contact for a broad resonance interaction initially grows linearly and follows the analytic expression derived in Ref.\;\cite{PhysRevA.91.013616} according to
\begin{equation}
\label{eq:ContactLinear}
n^{-4/3}\mathcal{C}_2(t) = \frac{128 \pi}{(6\pi^2)^{2/3}} \frac{t}{t_{\mathrm{n}}}.
\end{equation}
In addition, consistent with the derivation presented in App.\;\ref{subsec:broadresonancelimit}, the contact is observed to be proportional to the molecular condensate fraction in the broad resonance limit \cite{art:Werner2009}. \par
Considering resonances with large $k_{\mathrm{n}} R^*$ on the other hand, we recognize that the initial growth of the contact is gradually becoming less rapid as $k_{\mathrm{n}} R^*$ increases. In order to quantify the time it takes for the contact to evolve, we have computed the half-way time $\tau_c$. This time scale is defined as the time it takes to reach half the maximum value of the contact. 
The inset of Fig.\;\ref{fig:Contact} reveals how $\tau_c$ increases as a function of the resonance width. Similarly to the initial growth of the molecular lifetime, we observe the initial growth of the contact to evolve from linear to quadratic. In addition, the saturation time increases for larger values of $k_{\mathrm{n}} R^*$. We relate this to the increase in the mean transition time $t^*$ of the closed-channel molecules. The increase in $t^*$ slows down the formation of excitations from closed-channel molecules. Equation \ref{eq:ContactOriginal} then reveals how a delay in the saturation of the large momentum modes of the excited state fraction delays the saturation of the contact. As the large momentum modes of the excited state fraction saturate more rapidly than the smaller momentum modes \cite{art:makotyn,art:sykes}, we expect the contact in Fig.\;\ref{fig:Contact} to saturate before the excited state fraction in Fig.\;\ref{fig:Population}(b).
\subsection{\label{subsec:TheEmbeddedDimer} Dimer}
\begin{figure}[t!]
\centering
\includegraphics[width=8.6cm]{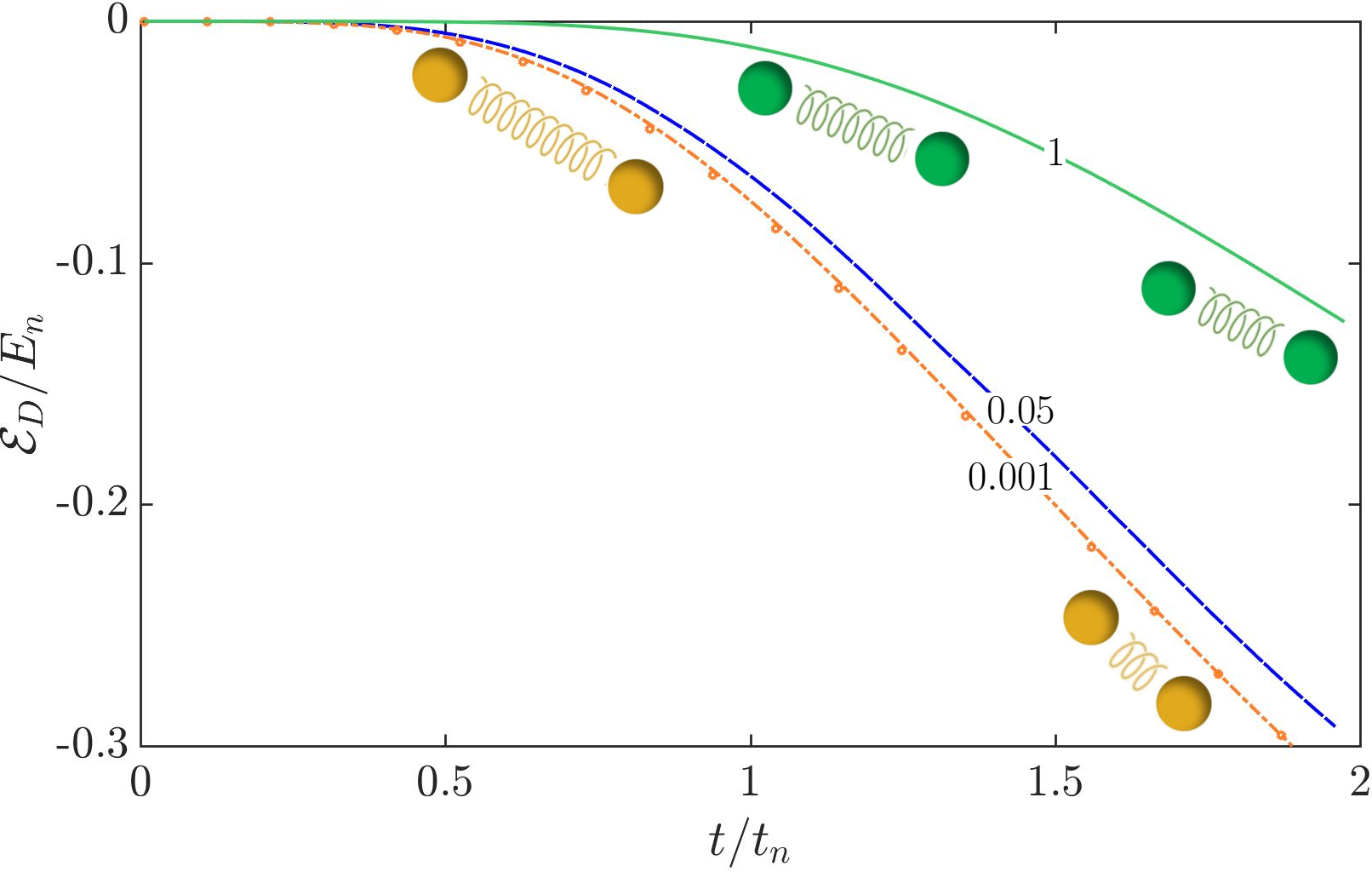}
\caption{Dimer binding energy rescaled with the Fermi energy as a function of the time spent at unitarity for three different values of the dimensionless resonance width $k_{\mathrm{n}}R^*$. The dimer shrinks progressively with increasing time at unitarity, as schematically illustrated by the cartoon. The dashed orange line is consistent with the single channel result (orange circles) as presented in Ref.\;\cite{art:colussimk}.}
\label{fig:DimerEnergy}
\end{figure}

\begin{figure}
      \begin{minipage}{.48\textwidth}
       \begin{tikzpicture}[>=latex]
\node at (0,0)
{\includegraphics[width=8.6cm]{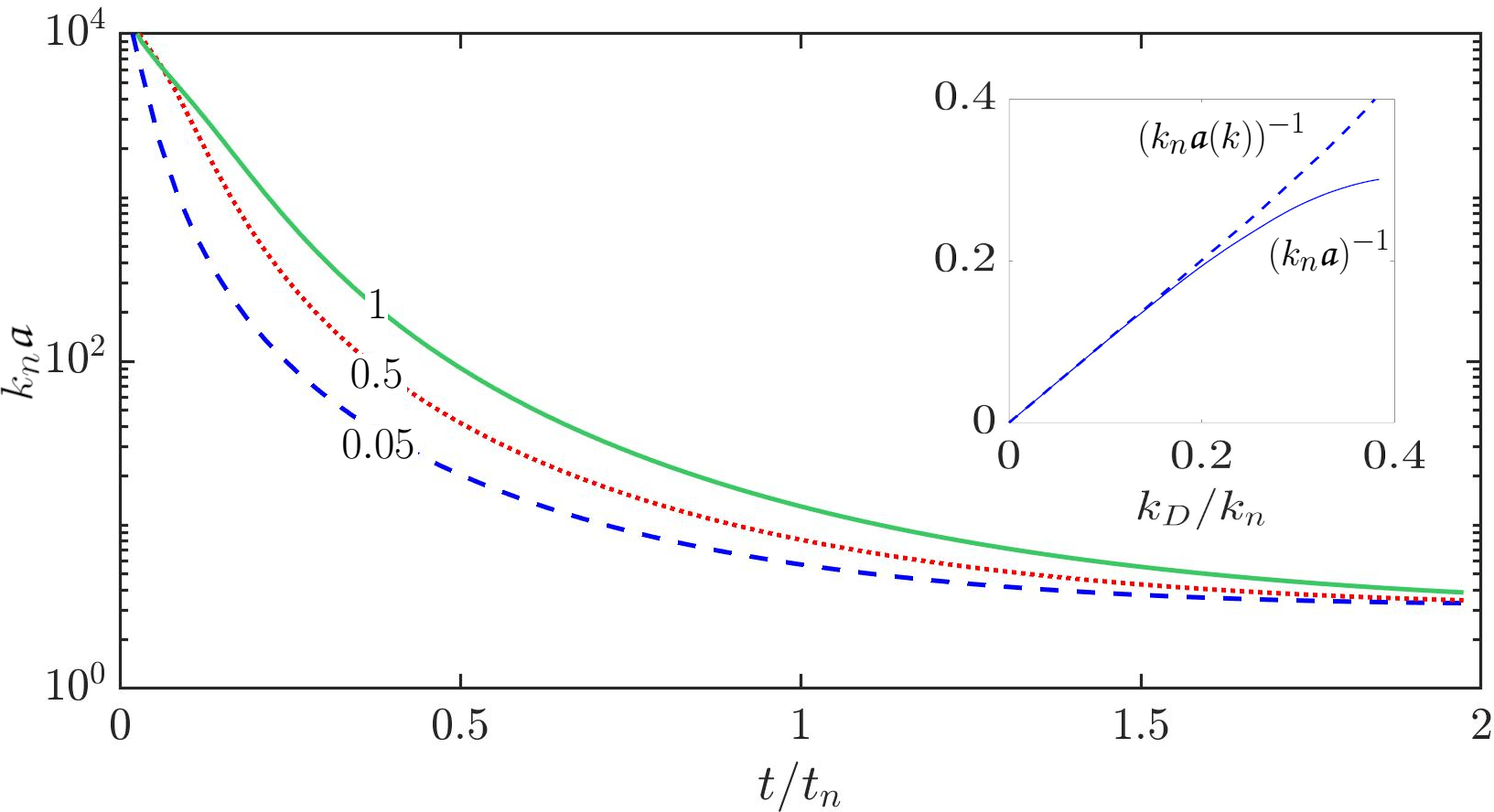}};
%\draw [->, thick] (0.5,1) -- (2.3,-0.1);
\node at (-4.1,-1.8) {(a)};
\end{tikzpicture}      
           \end{minipage}
  
      \begin{minipage}{.48\textwidth}
       \begin{tikzpicture}[>=latex]
\node at (0,0)
{\includegraphics[width=8.6cm]{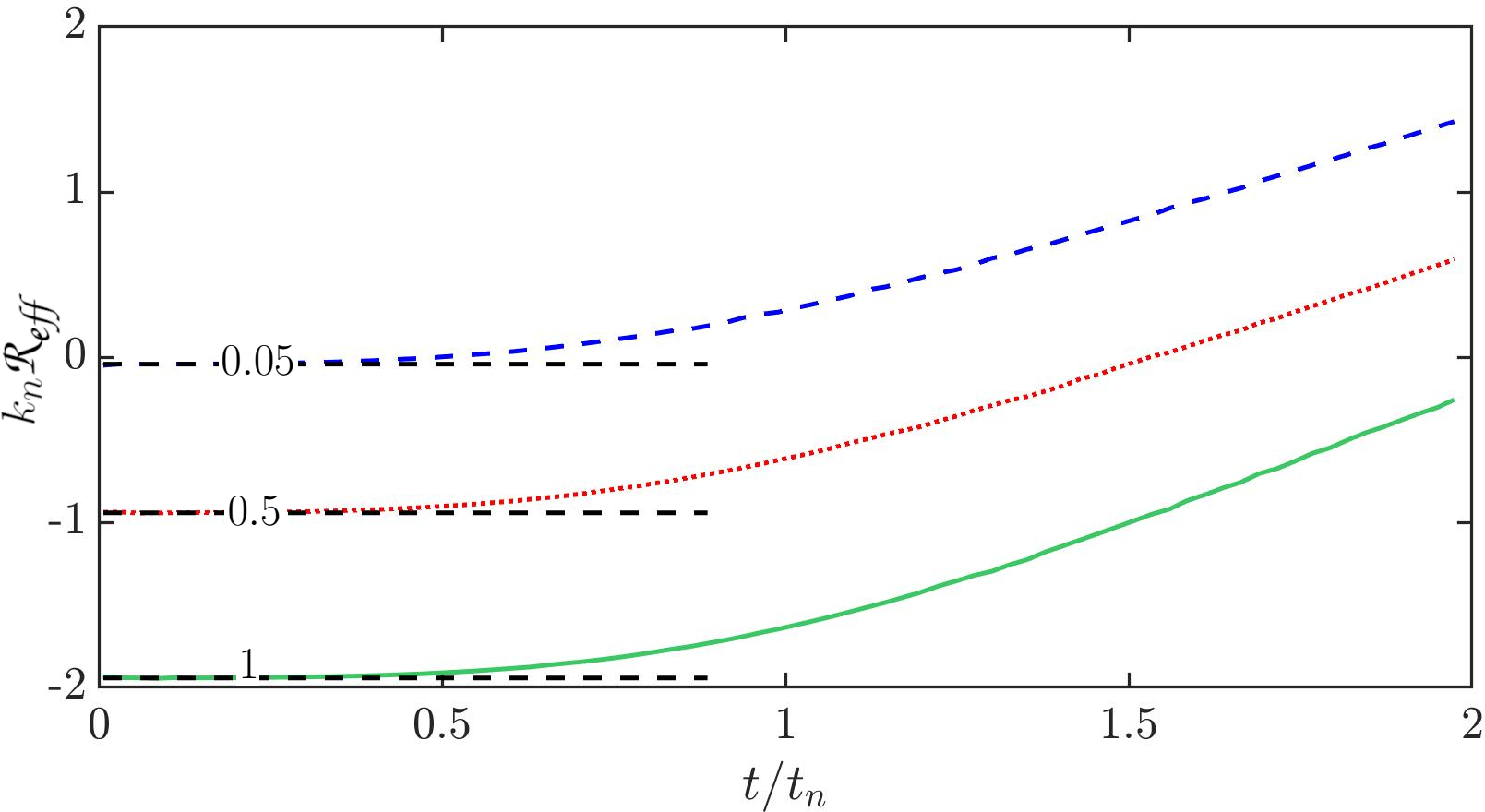}};
%\draw [->, thick] (0.5,1) -- (2.3,-0.1);
\node at (-4.1,-1.8) {(b)};
\end{tikzpicture} 
       \caption{The effective scattering length $\mathpzc{a}$ (5a) and the effective range $\mathpzc{R_{\mathrm{eff}}}$ (5b) as a function of the time spent at unitarity for various values of the dimensionless resonance width $k_{\mathrm{n}}R^*$. The inset in (a) presents the inverse of the scattering lengths $a$ and $a(k)$ as a function of the dimer wavenumber rescaled with the Fermi wavenumber.  Dashed black lines have been added to (b) in order to indicate the vacuum limit of the effective range as computed using Eq.\;\eqref{eq:effectiverangevacuum}
.} \label{fig:Scattering length}
      \end{minipage}
    \end{figure}

In this section we focus on how the evolution of the embedded dimer energy $\mathcal{E}_{\mathrm{D}}$ as investigated in the single channel limit in Refs. \cite{art:MusolinoPairFormation,art:colussimk} is affected by the gradual increase of $k_{\mathrm{n}}R^*$. This allows us to investigate how the size of the dimer and the universal scaling of the dimer energy with the scattering length is affected by the medium as well as the resonance width. The results of this analysis are presented in Figs.\;\ref{fig:DimerEnergy} and \ref{fig:Scattering length}. 
\par We observe that, contrary to vacuum two-body interactions, the dimer energy decreases towards more deeply bound values as the time spent at unitarity progresses. We relate this to the reduction of the magnitude of the embedded scattering length as presented in Fig.\;\ref{fig:Scattering length}(a), which indicates a decrease in the size of the embedded dimers as illustrated by the sketch in Fig.\;\ref{fig:DimerEnergy}.  \par
The sketch indicates that the localization is more rapid for increasing $k_{\mathrm{n}}R^*$, consistent with the production of more excitations for broader resonances, as discussed in Sec.\;\ref{subsec:populationdistribution}. 
Since these atoms Bose-enhance open-channel interactions, the change in the effectively experienced open channel potential $\hat{\mathcal{V}}_{\mathrm{eff}}$ is more drastic for broad resonances, such that the system is effectively pushed away from resonance more rapidly, corresponding to a more swift localization of the dimer. 
Besides commenting on the size of the embedded dimer, we notice that Fig.\;\ref{fig:Scattering length}(a) shows that, before the completion of the quench, the scattering length has already assumed finite values. This is a result of the, albeit limited, depletion of the initial state during the quench \cite{art:colussimk}. 
\par 
Similarly to two-body physics, we expect that the universal relation between the embedded scattering length and the dimer energy $\abs{\mathcal{E}_{\mathrm{D}}} \approx \hbar^2/(m\mathpzc{a}^{2})$ only holds for sufficiently large values of the scattering length. Furthermore, the correction to the universal quadratic relation becomes more important at larger values of the scattering length for increasingly narrow resonances. Therefore, in addition to analyzing the embedded scattering length, we present the evolution of the effective range $\mathpzc{R_{\mathrm{eff}}}$ as a function of the time spent at unitarity in Fig.\;\ref{fig:Scattering length}(b). 
Initially, the embedded effective range correctly reduces to the expected vacuum result as extracted from Eq.\;\eqref{eq:effectiverangevacuum}. However, as the time spent at unitarity progresses, the value of the embedded effective range starts to increase. \par 
Especially interesting is the observation that, at later times, the effective range assumes non-negligible values even for broad resonances and cannot be ignored. This effect is visualized in the inset of Fig.\;\ref{fig:Scattering length}, where the inverse scattering length $\mathpzc{a}^{-1}$ and the function $\mathpzc{a}(k)^{-1} = \mathpzc{a}^{-1} +\frac{1}{2}\mathpzc{R_{\mathrm{eff}}} k^2$ are plotted  versus the rescaled dimer wavenumber $k_{\mathrm{D}}/k_{\mathrm{n}} = \sqrt{\mathcal{E}_{\mathrm{D}}/2E_{\mathrm{n}}}$ and the curves are observed to differ for larger values of $k_{\mathrm{D}}/k_{\mathrm{n}}$. 
\subsection{\label{subsec:Z-parameter} Z-parameter}
\begin{figure}[t!]
\centering
\includegraphics[width=8.6cm]{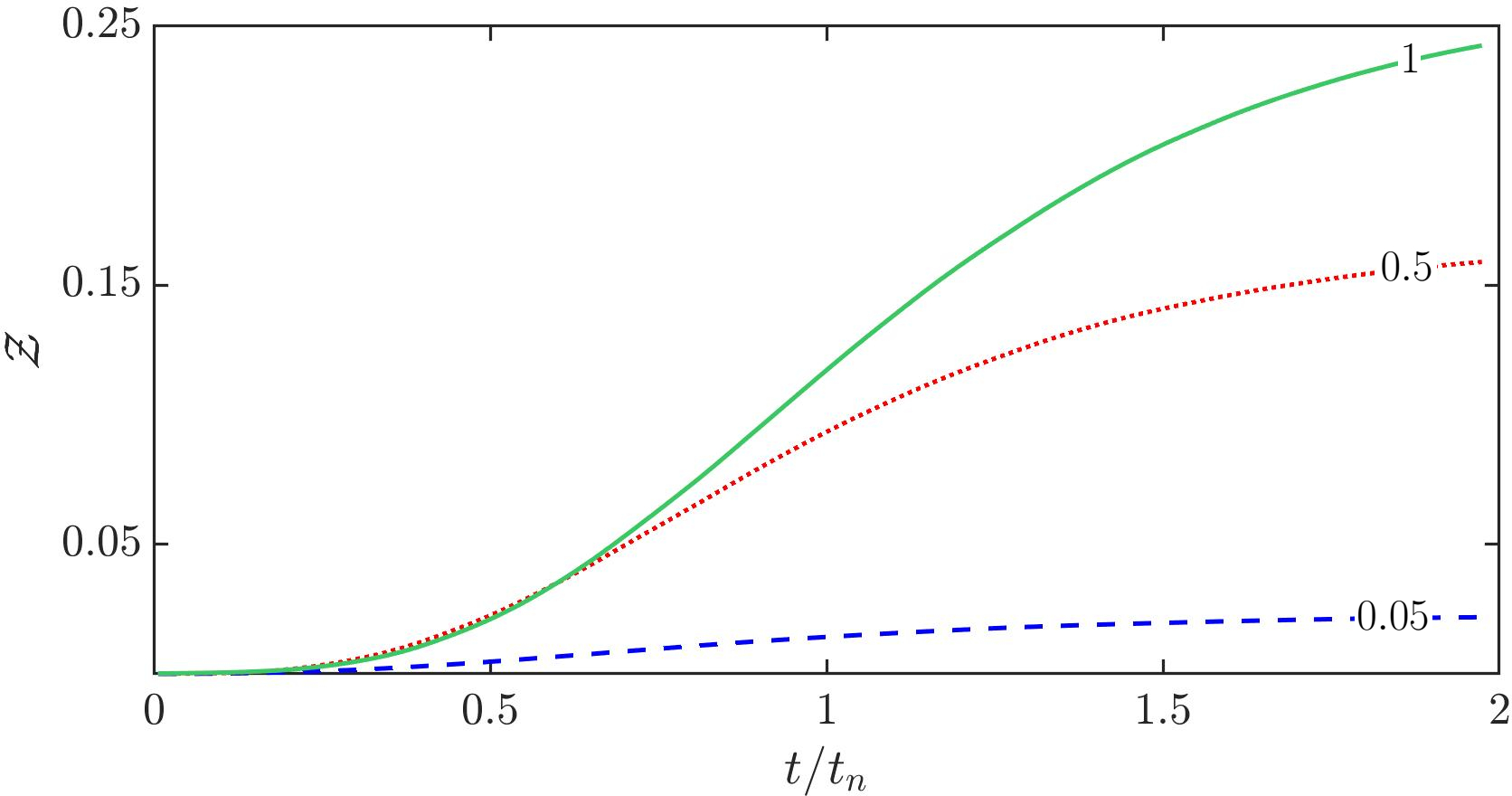}
\caption{The embedded Z-parameter (Eq.\;\eqref{eq:Zembedded}) as a function of the time spent at unitarity for various values of the dimensionless resonance width $k_{\mathrm{n}}R^*$.}
\label{fig:Z parameter}
\end{figure}

We now aim to analyze how the embedded dimer is distributed amongst the closed- and open-channel subspaces through the computation of the dynamical $\mathcal{Z}$-parameter. Consistent with Eq.\;\eqref{eq:Zvacuum} for a two-body system, our analysis of the $\mathcal{Z}$-parameter as presented in Fig.\;\ref{fig:Z parameter} reveals that initially $\mathcal{Z} = 0$ at unitarity.
We interpret this as being a result of the dimer extending over an infinite length scale at unitarity, such that the overlap of this infinite quantity with the localized closed-channel molecule vanishes. However, in agreement with the picture of the localization of the dimer as elaborated in Sec.\;\ref{subsec:TheEmbeddedDimer} and as predicted in Ref.\;\cite{art:stoof2009}, the $\mathcal{Z}$-parameter assumes non-zero values as the time spent at unitarity increases.   
Figure \ref{fig:Z parameter} shows that the finite value of the $\mathcal{Z}$ increases as a function of $k_{\mathrm{n}}R^*$, consistent with the dimer becoming progressively less open channel dominated for narrow resonances.    
\subsection{\label{subsec:Relative dimer-state occupation} Relative dimer state occupation}
\begin{figure}[t!]
\centering
\includegraphics[width=8.6cm]{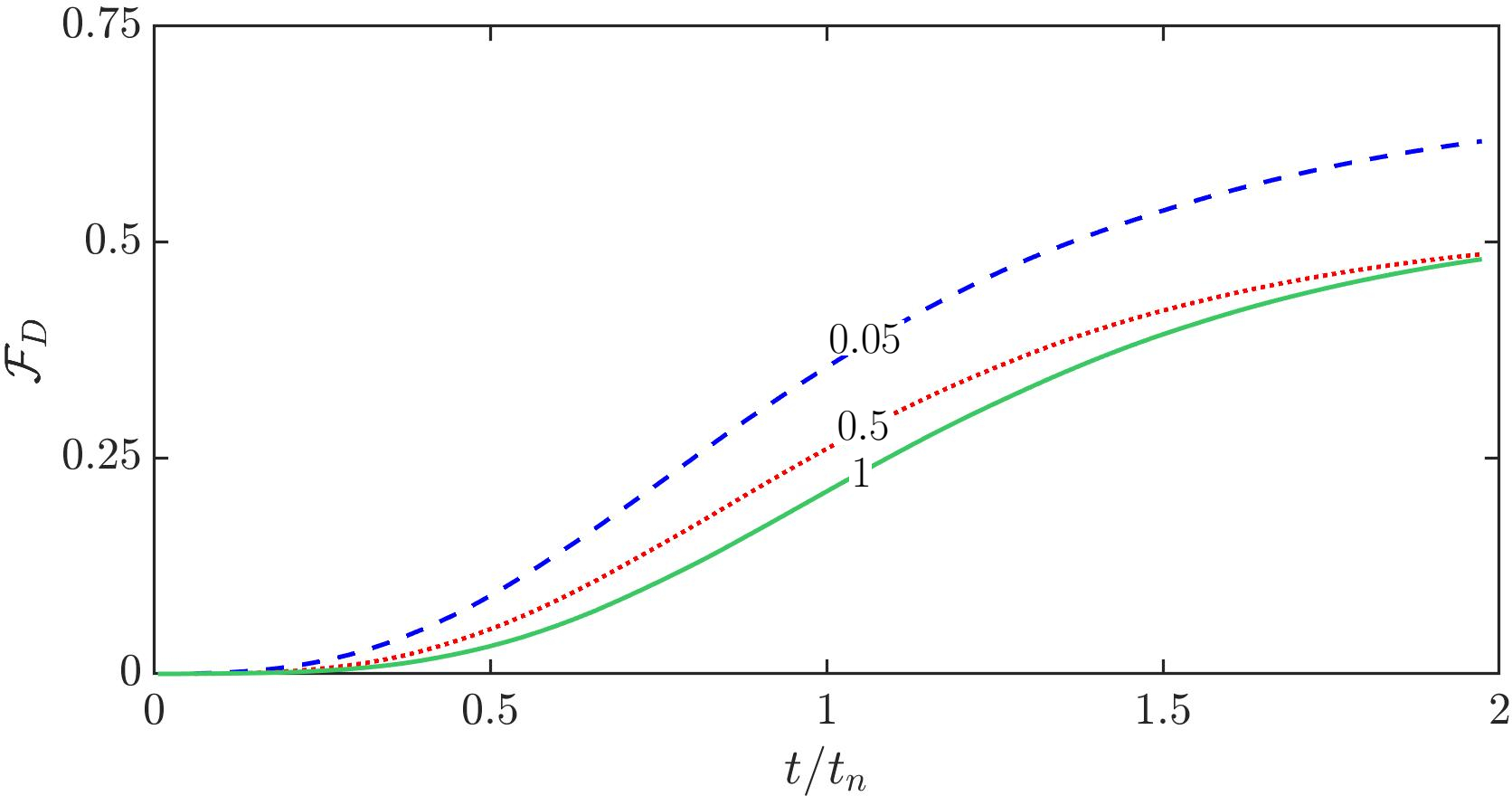}
\caption{The relative dimer state occupation (Eq.\;\eqref{eq:FparamDefinition}) for various values of the dimensionless resonance width $k_{\mathrm{n}}R^*$ as a function of the time spent at unitarity.}
\label{fig:F parameter}
\end{figure}

In the past two subsections, we have focused on the characteristics of the embedded dimer state. However, we are yet to quantify the relative importance of this dimer state with respect to the total pairing matrix. As outlined in Sec.\;\ref{subsec:embeddedFparameter}, this importance is gauged by the quantity $\mathcal{F}_{\mathrm{D}}$ as presented in Eq.\;\eqref{eq:FparamDefinition}. The evolution of this parameter for various values of the resonance width is presented Fig.\;\ref{fig:F parameter}. 
Consistent with the results found in Ref.\;\cite{art:kira}, the embedded dimer state dominates the quantum depletion in the case of relatively broad resonances, displaying evolution of an initially fast growth followed by a slower relaxation towards a finite value. 
Furthermore, Fig.\;\ref{fig:F parameter} reveals that this complicated dynamics is affected by the resonance width. In the simulated time-frame, the relatively broad-resonance case presents an upper-bound to the value of $\mathcal{F}_{\mathrm{D}}$. This means that the embedded dimer state is comparatively less relevant for pair correlations in narrow resonance systems. 
\par 
In order to gain an intuitive picture of this observation, we refer back to Fig.\;\ref{fig:DimerEnergy}, where we observe that dimers formed using narrow resonance interactions are less-deeply bound compared to dimers formed using broad resonance interactions. Consequently, narrow resonance interactions result in comparatively large dimers, which don't localize to sizes comparable to the Fermi scale during the considered time frame. As the pair correlation physics occurs on the level of the Fermi scales, the relatively extended dimers extend over a larger range than the typical two-body pairing physics. Hence, the observation that the dimer state is less relevant for narrower resonances, corresponding to lower values of the $\mathcal{F}_{\mathrm{D}}$. 
%Conclusion
\section{\label{sec:conclusion}Conclusion}
In this work, we employ a two-channel model with finite range pairwise interactions in order to study the effect of the resonance width on the dynamics of quenched unitary Bose gases. Using a two-channel model we vary the value of $k_{\mathrm{n}}R^*$, effectively changing the width of the resonance. This allows us to analyse the onset of non-universal effects. For increasing $k_{\mathrm{n}} R^*$ the lifetime of the closed channel molecules is increased relative to the Fermi scale, and the pairwise production of excitations in the open channel is decreased. Consequently the early-time growth of the dynamical two body contact becomes more gradual, transitioning from linear- to quadratic behaviors. In the limit $k_{\mathrm{n}} R^* \gg 1$, the early-time growth of the molecular condensate fraction also shifts from linear to quadratic and is set by the mean molecular lifetime $t_{*} = \sqrt{t_{\mathrm{n}}\tau}$. Here the system scales as combinations of resonance parameters and the density. Within the embedded few-body problem, we study the localization of dressed embedded dimers that are bound purely by many-body effects and are dressed by multichannel couplings. This analysis reveals their dominant role in the quantum depletion and how they become increasingly closed channel dominated as $k_{\mathrm{n}}R^*$ is increased.
\par Our analysis of the closed-channel contribution to the dressed embedded dimers and the pair condensate fraction opens up  additional experimental probes in the unitary regime \cite{PhysRevLett.95.020404,art:MusolinoPairFormation,art:hodby}. 
Furthermore, including third-order cumulants in the two-channel model along the lines of Refs.\;\cite{art:colussimk,PhysRevA.102.063314,musolino2021boseeinstein} represents an interesting prospect for future research, allowing for the investigation of Efimov physics, the triple condensate fraction and atom-molecule interactions in the unitary Bose gas for various values of the resonance width. 
\section{\label{sec:acknowledgements}Acknowledgments}
The authors thank Thomas Secker for supplying a full coupled-channels calculation data set in order to compute the momentum-space cut-off and acknowledge Paul Mestrom and Jing-Lun Li for useful discussions.  D.J.M.A.B, S.M. and S.J.J.M.F.K acknowledge financial support by the Netherlands Organisation for Scientific Research (NWO) under Grant No. 680-47-623. V.E.C. acknowledges financial support from Provincia Autonoma di Trento and the Italian MIUR under the PRIN2017 projectCEnTraL. 
\bibliographystyle{apsrev4-1}
\bibliography{biblio_paper}
\appendix
\vspace{0.5cm}
\section{\label{subsec:broadresonancelimit} Broad resonance limit}
In the vacuum broad resonance limit, the two-body interactions can be described by a single channel model, since the number of closed channel molecules remains small.
In the many-body environment, the broad resonance limit of Eqs.\;\eqref{eq:psia3}-\eqref{eq:kappa} can be obtained by eliminating the molecular wave function $\psi_m$ from this set of equations, meaning that $\partial_t \psi_m \approx 0$, such that
\begin{align}
\psi_m = -\frac{g}{2\nu}\psi_a^2 \zeta^*(0)-\frac{g}{2\nu}\sum_{\mathbf{k}\neq 0} \zeta^*(2\mathbf{k})\kappa_{\mathbf{k}}.
\end{align}
The reduced set of equations then corresponds to the single channel HFB equations as presented in Refs. \cite{art:colussimk,art:MusolinoPairFormation}, with  
\begin{equation}
\label{eq:v1ch}
v_{\mathrm{1ch}} = v-\frac{\abs{g}^2}{2\nu},
\end{equation}  
where $v_{\mathrm{1ch}}$ is the renormalized single channel potential interaction strength \cite{art:kokk,art:servaas_ram}. Additionally, 
in a single-channel model, the two-body transition matrix can be expressed as 
\begin{equation}
\label{eq:Tsinglechannel}
T_{\mathrm{1ch}} = v_{\mathrm{1ch}}\zeta(2\mathbf{k}) \braket{\zeta|\psi_{\mathrm{P}}^+}, 
\end{equation}
Apart from the difference in the potential strength, the previous expression corresponds to the uncoupled open channel part of the transition matrix as presented in Eq.\;\eqref{eq:Tvac}. In the broad resonance limit, Eq.\;\eqref{eq:Tsinglechannel} must match the expression for the two channel transition matrix as given in Eq.\;\eqref{eq:Tvac}, such that 
\begin{widetext}
\begin{equation}
\label{eq:AppendixEquationMatchingSingleToOther}
v_{\mathrm{1ch}} \zeta(2\mathbf{k}) \braket{\zeta|\psi_{P,1ch}^+} = v \zeta(2\mathbf{k}) \braket{\zeta|\psi_P^+} + \frac{\frac{g^2}{2}\abs{\braket{\zeta|\psi_P^+}}^2}{E-\nu-\frac{g^2}{2}\braket{\zeta|\hat{G}_P(E)|\zeta}},
\end{equation}
\end{widetext}
where $\braket{\zeta|\psi_P^+}$ can be expressed as 
\begin{equation}
\braket{\zeta|\psi_P^+} = \zeta(2 \mathbf{k})^* \left(1+\frac{v\xi(E)}{1+4\pi v \xi(E)}\right), 
\end{equation}
with 
\begin{equation}
\xi(E)= \frac{1}{(2\pi)^3} \int \abs{\zeta(2 \mathbf{k}')}^2 \frac{k'^2}{E-\hbar^2 k'^2/m} d k'.
\end{equation}
Furthermore, we can write the term $\braket{\zeta|\hat{G}_P(E)|\zeta}$ that appears in the denominator of Eq.\;\eqref{eq:AppendixEquationMatchingSingleToOther} as
\begin{equation}
\braket{\zeta|\hat{G}_P(E)|\zeta} = 4\pi \xi(E)\left(1+\frac{v\xi(E)}{1+4\pi v \xi(E)}\right).
\end{equation}
Substituting the expression for the potential strength $v$ as presented in Eq.\;\eqref{eq:vpotential} into Eq.\;\eqref{eq:AppendixEquationMatchingSingleToOther}, defining $v_{\mathrm{1ch}}$ analogously to $v$ and considering the low-energy limit, we can express the single channel scattering length $a_{\mathrm{1ch}}$ as
\begin{equation}
\label{eq:asingle}
a_{\mathrm{1ch}} = a_{\mathrm{bg}}-\frac{m}{4\pi \hbar^2}\frac{g_0^2}{2 \nu_0}.
\end{equation}
 Comparing Eq.\;\eqref{eq:asingle} to the definition of $a_{\mathrm{eff}}$ in Ref.\;\cite{PhysRevA.85.033616}, we recognize that the correction factor $g_c$ used in that work has been replaced by a factor $2$ in our definition of $a_{\mathrm{1ch}}$.  
There, the correction factor $g_c$ was set to a value of 1.816 in order to match the binding energy of the contact potential model as closely as possible. A similar parametrization and correction factor was found in Ref.\;\cite{art:servaas_ram}. We recognize that, using the separable potential model, our analytically obtained factor of $2$ is in relatively close agreement to the calibrated correction factors used in these works.

Extending the broad-resonance limit analysis to the two-body contact, we realize that this limit allows for the computation of the contact using the adiabatic sweep theorem \cite{TAN20082971,art:Werner2009} 
\begin{align}
\mathcal{C}_2 = \frac{4\pi m}{\hbar^2}\frac{d E}{d(-1/a)}.
\end{align}
Applying the dispersive relation between the scattering length and the magnetic field as presented in Eq.\;\eqref{eq:ScatteringFeshbach}, the previous expression can be rewritten into the form
\begin{align}
\mathcal{C}_2 = \frac{4\pi m}{\delta\mu R^* (1-a_{\mathrm{bg}}/a)^2}\frac{d E}{d B}.
\end{align}
Since only the bare bound-state energy of the closed-channel molecule depends on the magnetic field, the application of the Helmann-Feynmann theorem \cite{art:Werner2009,PhysRevA.78.053606} reveals that the two-body contact can be expressed as \cite{art:Werner2009,book:Castin2012}
\begin{align}
\mathcal{C}_2 = \frac{4\pi \abs{\psi_m}^2}{R^*}\left(1-\frac{a_{\mathrm{bg}}}{a}\right)^{-2}.
\end{align}
The previous expression implies that the two-body contact scales linearly with the molecular condensate fraction in the broad-resonance limit. 
\section{\label{subsec:RenormalizationPairCondensate} Renormalization of the pair condensate}
\begin{figure}[h!]
\centering
\includegraphics[width=8.6cm]{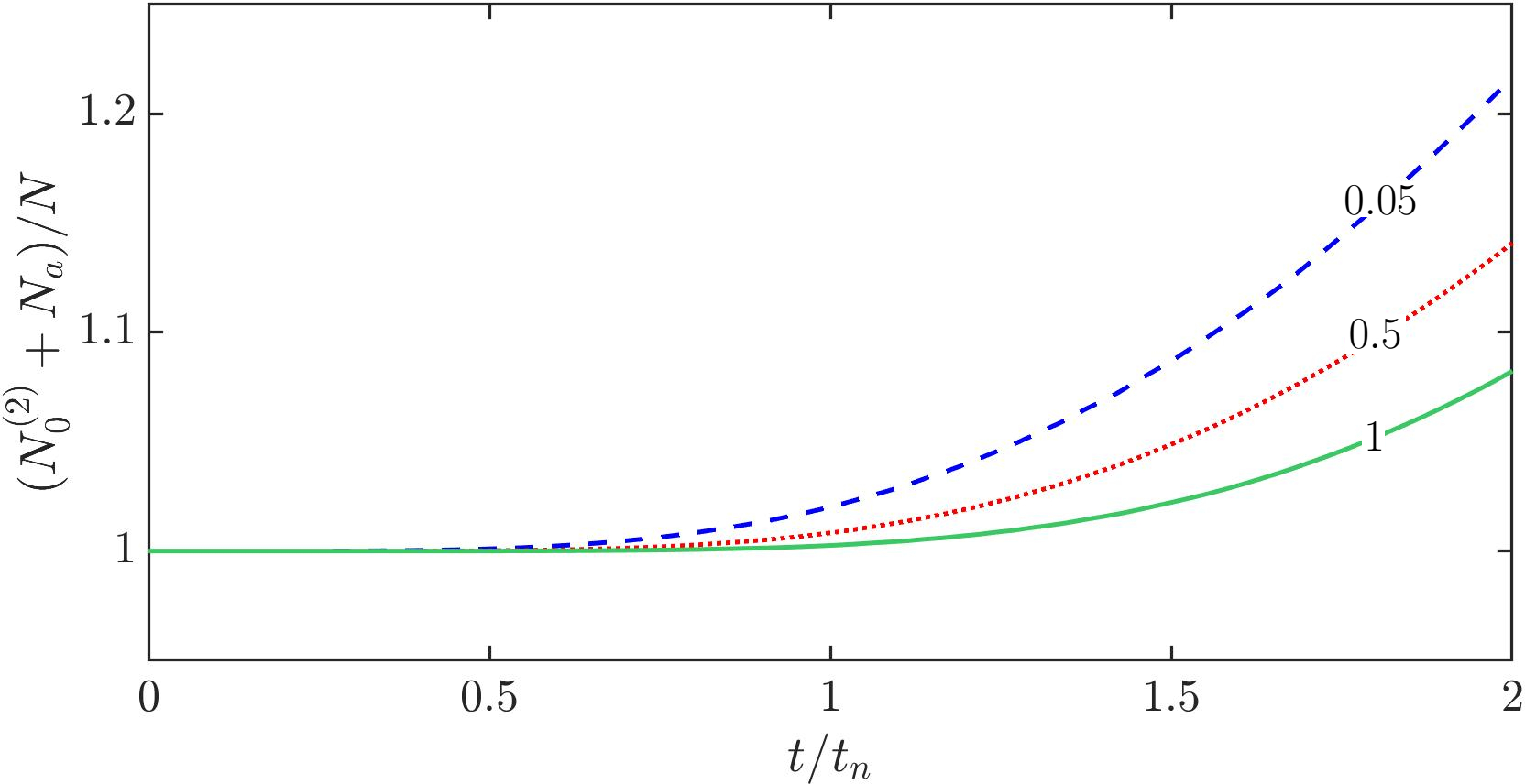}
\caption{The unrenormalized pair fraction $N_0^{(2)}$ added to the atomic condensate fraction as a function of the time $t/t_{\mathrm{n}}$ spent at unitarity.}
\label{fig:unrenormalized}
\end{figure}

As outlined in Sec.\;\ref{subsec:PairCondensation}, we cannot directly relate the eigenvalue $N_0^{(2)}$ as presented in Eq.\;\eqref{eq:N02Eigenvalue} to the number of condensed pairs due to the Bose enhancement that alters the bosonic commutation relation and results in an over counting of the number of pairs. As presented in Fig.\;\ref{fig:unrenormalized}, the unrenormalized number of pairs exceeds the number of particles available for pairing $N-N_a$, such that $N_0^{(2)}+N_a >N$. The overcounting of bosons at early times is larger for broader resonances due to the increased value of the open-channel pairing matrix $\kappa_{\mathbf{\mathrm{k}}}$.
In order to retrieve the number of condensed pairs and the correct bosonic commutation relations, the eigenvalue $N_0^{(2)}$ has to be renormalized. To this extend we define the following momentum space composite operator that annihilates a condensed pair 
\begin{align}
\label{eq:d0}
\hat{d}_0 = \frac{1}{\sqrt{2}}\sum_{k \neq 0} \begin{bmatrix} \hat{a}_{k}\hat{a}_{-k} \varphi_{P,0}(k) \\ \sqrt{2} \phi(k) \hat{b}_0 \varphi_{Q,0}(k) \end{bmatrix},
\end{align} 
such that $\braket{d_0^{\dagger}d_0} = N_0^{(2)}/2$. 
By calculating the expectation value of the commutator $[\hat{d}_0,\hat{d}^{\dagger}_0]$, we can compute the overcounting factor required for the renormalization. Using Eq.\;\eqref{eq:d0} we find that 
\begin{align}
\label{eq:Commutator1}
\left[\hat{d}_0,\hat{d}^{\dagger}_0\right] &= \frac{1}{2} \sum_{k,k' \neq 0} \left(\varphi_{P,0}(k)\varphi_{P,0}(k') [\hat{a}_k \hat{a}_{-k},\hat{a}^{\dagger}_{k'} \hat{a}^{\dagger}_{-k'}] \right. \notag \\ 
& \left.+ 2 \varphi_{Q,0}(k)\varphi_{Q,0}(k') \phi(k) \phi(k') [\hat{b}_0,\hat{b}^{\dagger}_0]\right),
\end{align}
where we have used that the open and closed channel operators commute. 
Since both operators satsify the bosonic commutation relation and using the orthonormality of the bare closed channel wave function, the expectation value of Eq.\;\eqref{eq:Commutator1} can be expressed as  
\begin{align}
\label{eq:CommutatorExpectation1}
\left\langle\left[\hat{d}_0,\hat{d}^{\dagger}_0\right]\right\rangle = \sum_{k \neq 0} \frac{\abs{\braket{\hat{a}_k \hat{a}_{-k}}}^2}{N_0^{(2)}}(1+2\hat{a}_k^{\dagger}\hat{a}_k)+2\frac{\abs{\braket{\hat{b}_0}}^2}{N_0^{(2)}}, 
\end{align}
where we have used the momentum space analogue of the pair wave function presented in Eq.\;\eqref{eq:PairWaveFunctionr} in order to express the eigenvector components in terms of the expectation value of the cumulants introduced in Sec.\;\ref{subsec:manybodyeq}. 
Recognizing the factor $N_0^{(2)}$ as defined in Eq.\;\eqref{eq:N02Eigenvalue}, we can rewrite Eq.\;\eqref{eq:CommutationExpectation} as 
\begin{align}
\label{eq:CommutationExpectation}
\left\langle\left[\hat{d}_0,\hat{d}^{\dagger}_0\right]\right\rangle = 1 + \frac{2}{N_0^{(2)}}\sum_{k \neq 0} \abs{\braket{\hat{a}_k \hat{a}_{-k}}}^2 \hat{a}_k^{\dagger}\hat{a}_k.
\end{align}
Since we neglect the presence of background molecules in the closed channel, Eq.\;\eqref{eq:CommutationExpectation} is identical to its single channel analogue presented in Ref.\;\cite{musolino2021boseeinstein}, despite the presence of the closed channel molecular fraction in the definition of $N_0^{(2)}$ presented in Eq.\;\eqref{eq:N02Eigenvalue}.
In order to retreive the desired bosonic commutation for the pair condensate, Eq.\;\eqref{eq:CommutationExpectation} inspires us to define the following renormalized analogue of the pair operator presented in Eq.\;\eqref{eq:d0}
\begin{align}
\hat{D}_0 = \frac{\hat{d}_0}{\sqrt{1 + \left[2/N_0^{(2)}\right]\sum_{k \neq 0} \abs{\braket{a_k a_{-k}}}^2 a_k^{\dagger}a_k}}
\end{align}
Using this renormalized operator, we can now compute the number of condensed pairs using $\braket{D_0^{\dagger}D_0}$ for various values of the resonance width (which in the many-body system can be parametrized in terms of $k_n R^*$). 
\section{\label{subsec:GeneralizedTanContact} Relating the contact to cumulants}
In order to relate Eq.\;\eqref{eq:CorrelationGeneralized} to the set of cumulants introduced in Sec.\;\ref{subsec:manybodyeq} such that we can compute its value as a function of the time spent at unitarity,  we consider the zero energy limit of Eq.\;\eqref{eq:psim}, finding that 
\begin{align}
\hat{\phi}_{\mathrm{Q}} = -\frac{g^2}{2\nu} \hat{\psi}_{\mathrm{P}}\hat{\psi}_{\mathrm{P}},
\end{align}
where we have used $\braket{\hat{\psi}_P} = \zeta(0)\psi_a^2+\sum_{\mathbf{k}} \zeta(2\mathbf{k}) \kappa_{\mathbf{k}}$ and $\braket{\hat{\phi}_{\mathrm{Q}}} = \psi_m$. 
Analogous to Ref.\;\cite{PhysRevA.78.053606} we can then define the following compound operator $\hat{\Phi}$
\begin{align}
\hat{\Phi} = v\hat{\psi}_{\mathrm{P}}\hat{\psi}_{\mathrm{P}} + g \hat{\phi}_{\mathrm{Q}},
\end{align}
such that we can obtain the following  expression for the generalized Tan contact
\begin{align}
\mathcal{C}_2 = \braket{\hat{\Phi}^{\dagger}\hat{\Phi}}.
\end{align}
In terms of our set of cumulants the previous expression shows that the two-channel contact can be computed as 
\begin{widetext}
\begin{equation} 
\label{eq:ContactGeneralizedFinal}
\begin{split}
\mathcal{C}_2(t) = &\frac{m^2 v^2}{\hbar^4} \left(\abs{\psi_a}^4 \abs{\zeta(0)}^2+4\abs{\psi_a}^2\sum_{\mathbf{k}}\abs{\zeta(\mathbf{k})}^2\rho_{\mathbf{k}} +2\sum_{\mathbf{k},\mathbf{k}'}\abs{\zeta(\mathbf{k}-\mathbf{k}')}^2 \rho_{\mathbf{k}}\rho_{\mathbf{k}'}+\sum_{\mathbf{k}}\left[ \zeta(2\mathbf{k})\zeta^*(0) \kappa_{\mathbf{k}}^* \psi_a^2 +\text{h.c.} \right] \right.\\  
& \left. +\abs{\sum_{\mathbf{k}}\zeta^*(\mathbf{k})\kappa_{\mathbf{k}}}^2  + \frac{1}{\sqrt{2\pi a_{\mathrm{bg}}^2 R^*}}\left[(\psi^{\dagger}_a)^2\psi_m \zeta(0)+\sum_{\mathbf{k}} \zeta(2\mathbf{k})\kappa_{\mathbf{k}}^*\psi_m + \text{h.c.} \right] +\frac{1}{2 \pi a_{\mathrm{bg}}^2 R^*} \abs{\psi_m}^2\right).
\end{split}
\end{equation}
\end{widetext}
Equation \eqref{eq:ContactGeneralizedFinal} is used to obtain the results as presented in Sec.\;\ref{subsec:TwoBodyContact}.
\end{document}